\documentclass[sigconf, nonacm]{acmart}
\usepackage{multirow} 
\usepackage{hyperref}
\usepackage{bm}
\setcopyright{none} 
\newcommand{\hl}[1]{#1}
\usepackage{balance}
\providecommand{\hlsec}[1]{#1}

\acmConference[]{}{}{}

\AtBeginDocument{%
  }

\DeclareMathOperator*{\argmin}{arg\,min}

\begin{document}

\title[Heterogeneous Interaction Network Analysis (HINA)]{Heterogeneous Interaction Network Analysis (HINA): A New Learning Analytics Approach for Modelling, Analyzing, and Visualizing Complex Interactions in Learning Processes}

\author{Shihui Feng}
\email{shihuife@hku.hk}
\affiliation{%
  \institution{University of Hong Kong}
  \city{Hong Kong}
  \country{China}
}
\authornote{Corresponding author} 

\author{Baiyue He}
\email{baiyue.he@connect.hku.hk}
\affiliation{%
  \institution{University of Hong Kong}
  \city{Hong Kong}
  \country{China}
}

\author{Dragan Gasevic}
\email{dragan.gasevic@monash.edu}
\affiliation{%
  \institution{Monash University}
  \city{Melbourne}
  \country{Australia}
}

\author{Alec Kirkley}
\email{akirkley@hku.hk}
\affiliation{%
 \institution{University of Hong Kong}
 \city{Hong Kong}
 \country{China}
}

\begin{abstract}
Existing learning analytics approaches, which often model learning processes as sequences of learner actions or homogeneous relationships, are limited in capturing the distributed, multi-typed interactions in contemporary learning environments. To address this, we propose Heterogeneous Interaction Network Analysis (HINA), a multi-level learning analytics framework for modelling interactions and associations across diverse entities (e.g., learners, behaviours, and AI agents) in learning processes. Grounded in network science principles, HINA integrates a multi-level analytical framework that analyzes individual process metrics (node-level), prominent associations (dyad-level), and latent clusters (meso-level) to address questions about how different elements in a learning environment interact and co-influence each other. In this paper, we first detail the theoretical and mathematical foundations of HINA for individual, dyadic, and meso-level analysis. We then demonstrate HINA's utility through a case study on AI-assisted small-group collaborative learning, revealing students' interaction profiles with peers versus AI, distinct engagement patterns that emerge from these interactions, and specific types of learning behaviors (e.g., asking questions) directed to AI versus peers. HINA contributes a novel and unified analytical framework that enables researchers to quantify individual-level processes, identify significant associations, and uncover meaningful clusters within a single workflow. Accompanied by a dedicated web tool, HINA provides a complete workflow for supporting process-based assessment, and enables a new theoretical lens for modeling and understanding complex and mediated learning processes. 
\end{abstract}


\maketitle

\section{Introduction}

Learning is fundamentally an interactive process, emerging from the diverse and multifaceted engagement a learner has with their designed learning environment---including peers, instructors, digital tools, and pedagogical artefacts \citep{bandura1977social,chi2009active, knight2020implementing}. Constructivist learning theory underscores that knowledge is actively constructed through these various interactions \citep{vygotsky1978mind, piaget2013construction}, making the detailed study of learning processes essential to understanding how and why learning occurs. These processes are characterized not only by homogeneous interactions (e.g., student-to-student), but, critically, by a web of heterogeneous interactions among different entity types, such as a student's use of a specific cognitive strategy, their contributions to learning artefacts, or their engagement with an AI tool's function \citep{goodyear2019analysis, Feng2025JLA, gibson2023learning}.

Understanding how learning occurs requires us to model the heterogeneous interactions that fundamentally constitute learning processes \citep{suthers2017multilevel,goodyear2019analysis,poquet2022cacophony,castaneda2024relational}. Existing learning analytics approaches (e.g. sequential analysis) are not designed to represent this complexity \citep{knight2017time,jovanovic2017learning,lamsa2020potential,molenaar2022towards}. They often model learning processes as temporally ordered sequences drawn from a single type of data such as log-file events and coded behaviors \citep{blikstein2016multimodal}, which limits their ability to capture and disentangle the distributed, multi-faceted nature of interactions in contemporary learning environments. Consequently, they cannot represent or analyze the rich, structured relationships---such as those between a student’s strategic action, a specific function of a digital tool, and a pedagogical artefact---that constitute the learning process \citep{chen2022networks,poquet2022cacophony,shi2016survey}. Crucially, there is a lack of an integrated analytical framework that provides quantifiable metrics, pattern discovery algorithms, and visualization for analyzing heterogeneous interactions in learning processes. The Heterogeneous Interaction Network Analysis (HINA) framework is developed to address these critical methodological gaps.

This study introduces Heterogeneous Interaction Network Analysis (HINA), a learning analytics framework designed to capture the complex structure of learning processes involving diverse entities (e.g. students, artefacts, AI tools, coded behaviors). Building on network science \citep{newman2018networks}, HINA operates as a domain-specific framework for learning analytics by explicitly modeling multi-typed entities (e.g., students, behaviors, artifacts, AI agents) and the semantic relationships among them. HINA transforms learning process data into heterogeneous interaction networks (HINs) and integrates a specialized analytical framework that includes node-level metrics, dyad-level edge pruning, and meso-level non-parametric clustering. HINA's multi-level analytical framework addresses the gap for quantifying, clustering, and visualizing heterogeneous interactions in learning processes. It supports both exploratory discovery of latent patterns and theory-based hypothesis testing. For instance, HINA enables researchers to identify how specific digital tools drive behaviors or how strategies emerge from interactions with learning artifacts. HINA thus provides a systematic approach for modeling heterogeneous interactions in learning processes and extracting meaningful theoretical and pedagogical insights from these complex structures.

Beyond methodological foundations, the choice of the term `Heterogeneous Interaction Network' is rooted in the theoretical requirements of learning analytics \citep{gasevic2015let}. While the cognate term `Heterogeneous Information Network' in computer science describes general relational dependencies between multi-typed entities \citep{shi2016survey}, it lacks a specific disciplinary focus. In contrast, Heterogeneous Interaction Networks emphasize the socio-cognitive interaction mechanisms that drive the learning process \citep{vygotsky1978mind}. By naming the framework HINA and identifying its foundational structure as Heterogeneous Interaction Networks (HINs), we explicitly prioritize the modeling of complex mediated interactions, such as those between learners, AI agents, and pedagogical artifacts. This terminology highlights the analytical focus of HINA, shifting from a general data-science perspective to an educational perspective centered on the dynamic and distributed nature of interactions in modern learning environments.

In this paper, we first present the conceptual, theoretical, and methodological foundations of the HINA framework. We then demonstrate its application through a case study in which small student groups collaborate with a generative AI chatbot, and report what HINA shows about human–AI interaction in that setting. Finally, we discuss the broader implications of HINA for modeling learning processes across diverse contexts and conclude with directions for future research, covering both the theoretical questions HINA can address about human–AI interaction and other learning contexts, as well as the key directions for HINA's continued development.


\section{Background}

\subsection {Heterogeneous Interaction Network Analysis (HINA): Concepts and Definitions}

Within the HINA framework, the fundamental entities of interest in the learning process data (e.g., students, learning objects, coded behaviors) are represented as nodes in a network. \textbf{Heterogeneous interactions} among these nodes are defined as relational ties formed between nodes of different types during the learning process. For example, a student interacting with a learning object is an instance of a heterogeneous interaction. A \textbf{Heterogeneous Interaction Network (HIN)} is a flexible graph structure designed to model the interactions between different types of entities in a learning process.

A HIN takes the form of a weighted bipartite graph among a specified pair of distinct entity types to model the designated heterogeneous interactions. For example, if one aims to study how students interact with learning objects, they can construct a Student-Object HIN. \hl{While HINA utilizes the mathematical structure of weighted bipartite graphs, it departs from standard bipartite models through the use of composite nodes. In HINA, a node can represent a higher-order semantic construct—such as a ``Tool-Action'' pairing. This data transformation allows HINA to project triadic relationships into a bipartite space, enabling the analysis of complex, multi-way interactions (e.g., Which students use which AI tools for which specific cognitive processes?).} We can increase granularity by considering composite entities formed by combinations of other entities. For example, a particular object being interacted with using a particular coded interaction behavior can be represented using a composite entity of the form \textit{(Object, Behavior Code)}. Thus, if one aims to understand how different learning objects and coded interaction behaviors are collectively engaged with by individual students, they can construct a HIN consisting of students, objects, and behavior codes. This models the triadic relationship of students engaging with specific objects via specific behaviors. Through this flexible structure, the HINA framework enables researchers to model diverse learning contexts by specifying the relevant heterogeneous interactions central to their research questions. Formally, we can define a HIN as a graph $G = (V_1, V_2, E, \omega)$, where: 
\begin{itemize} 

\item $V_1$ and $V_2$ are the two distinct node sets of interest. In the above examples, $V_1$ consists of all students, and $V_2$ would either consist of all learning objects (in the Student-Object HIN) or all learning object-coded behavior pairs (in the Student-(Object, Behavior Code) HIN). The ability to encode interactions with composite nodes formed from multiple entities simultaneously (e.g., actor-tool-action) provides HINs with a unique capacity for modeling the complex, mediated interactions inherent in learning processes. \\

\item $E\subseteq V_1\times V_2$ is the set of heterogeneous interactions, consisting of all node pairs $(i,j)$ for which node $i\in V_1$ and node $j\in V_2$ interact during the learning process. \\

\item $\omega: E \rightarrow \mathbb{N}$ is a function that assigns a positive integer-valued weight to each edge $(i,j)\in E$ quantifying the strength, frequency, or intensity of the interaction, as required by the specific research context. For notational brevity, we let $w_{ij}=\omega((i,j))$ denote the weight of the interaction $(i,j)$ according to the weight function of interest.

\end{itemize}

\subsection {Theoretical Underpinnings}

The HINA framework is theoretically grounded in constructivism theory and actor-network theory (ANT) \citep{vygotsky1978mind, piaget2013construction, latour1996actor}. From a constructivist viewpoint, learning is not the passive reception of information but an active process of meaning-making, where knowledge is constructed through the learner's dynamic interactions with their environment \citep{vygotsky1978mind, piaget2013construction}. HINA operationalizes this core tenet by formally capturing various interactions---among students, artefacts, and conceptual constructs along with other entities---to model learning processes. Interactions in learning processes are not limited to human-human social interactions, echoing ANT \citep{latour1996actor, callon1999actor} which provides a rethinking of agency. ANT posits that agency is a human attribute as well as emerges from networks of heterogeneous ``actants'', which encompasses humans, technologies, tools, and abstract concepts \citep{callon1999actor}. This principle directly informs the structure of HINs in HINA, where the various entities involved in learning processes (including students, learning objects, and coded behaviors) can all be treated as distinct types of nodes. By modelling the heterogeneous relationships among different types of entities, HINA moves beyond human-human social interactions to trace how knowledge and agency are co-constructed and distributed through students' interactions with the various entities in a designed learning environment.

HINA's theoretical commitment to heterogeneous interactions is a necessary response to the increasing complexity of digital and hybrid learning ecosystems, particularly those involving human-AI interaction. Traditional social network analyses often ignore non-human elements, such as pedagogical artifacts or specific cognitive behaviors. However, in modern digitally enhanced learning environments, a student’s interaction with an AI agent or a specific digital tool is both a technical event \citep{castaneda2024relational} and a constitutive element of the learning process itself. This is especially relevant for human-AI interaction, where the AI agent is an active actant that responds, adapts, and co-produces knowledge with the student. The HINA modelling structure allows us to distinguish both whom a student interacts with as well as what is exchanged, which offers the critical analytical basis for understanding how agency is distributed between learner and machine. By modeling these relationships as heterogeneous interaction networks, HINA preserves the semantic specificity of the learning trace. Heterogeneity is not just a structural feature in the HINA framework, but also a theoretical perspective for modelling and understanding interactions in complex and mediated learning. 

\setlength{\parskip}{0pt plus 4.5pt}

\subsection{Review of Learning Analytics Methods}\label{sec:review}

Learning analytics research has employed methodologies from other fields and developed its own for modelling and understanding learning processes. These methods can be broadly categorised according to their core analytical focus. A primary strand of research focuses on temporal and sequential patterns. Techniques such as process mining \citep{van2012process}, sequence mining \citep{jovanovic2017learning}, lag sequential analysis (LSA) \citep{lamsa2020potential}, and Hidden Markov Models (HMMs) \citep{jeong2010analysis} treat learning as a sequence of events or states \citep{baker2016educational, romero2020educational}, and are effective for identifying common pathways and transitions \citep{taylor2024quantifying,lamsa2020potential,jeong2010analysis}.

Another major category of analytical approaches employs clustering, dimensionality reduction, and pattern mining to group learners or features based on similar profiles derived from aggregated metrics (e.g., behavior frequencies, performance scores). Common methods, such as K-means clustering \citep{kodinariya2013review}, factor analysis \citep{lawley1962factor}, and association rule mining \citep{liu1998integrating}, identify archetypes of engagement, latent subgroups, and correlated feature sets \citep{hong2020latent, bharara2018application,whitelock2019student}. However, a main constraint for many of the clustering methods is their sensitivity to the selection of input features and a priori specification of algorithmic parameters (e.g., the number of clusters)\hl{. Existing methods offer no principled way to decide whether a grouping is warranted at all}. 

A third analytical category is network analysis, which moves beyond aggregated or sequential views to focus explicitly on modeling the relational dependencies among entities within a learning system. Network analysis models systems as sets of nodes (representing entities) connected by edges (representing relationships), which makes relational structure directly analyzable \citep{newman2018networks}. In learning analytics, this approach is easily applicable because learning is increasingly understood as a social and cognitive process embedded within networks of interactions, resources, and ideas \citep{vygotsky1978mind, poquet2022cacophony}. Early and well-established applications of network analysis in the field have primarily focused on social network analysis to map the structure of social interactions in online forums or collaborative environments \citep{saqr2019role,chen2018fostering}. These studies productively use metrics like centrality and density or inference methods such as stochastic actor-oriented models (SAOMs) or exponential random graph models (ERGMs) to analyze students' structural positions \citep{wang2024will,wu2021research}, interaction dynamics, as well as relate these patterns to learners' discourse and their prestige in the network \citep{zou2021exploring}.  Recognizing that social interactions represent only one dimension of the learning process, learning analytic research has moved beyond student-to-student social interactions in online forums to investigate the relationships between students and posts \citep{poquet2020forum}. By comparing observed network metrics against randomized null models, the study \citep{poquet2020forum} illustrated that students' structural positions in forums are often highly related to their posting behaviors, which shows the utility of bipartite structures for modeling student engagement in learning processes. 

Within learning analytics, several network-based methods have been developed. Epistemic Network Analysis (ENA) was developed to model the structure of knowledge constructed in discourse. ENA moves beyond social ties to model the co-occurrence of coded concepts within defined units of dialogue \citep{shaffer2016tutorial}. By constructing a network where nodes are concepts and edges represent their co-occurrence within a defined time-window, ENA \hl{projects the co-occurrence networks of all units of analysis into a shared low dimensional space, in which both the code nodes and each unit's network centroid have positions. The strength of connections in a student's or group's cognitive framework, and differences between groups, are then read from these positions} \citep{shaffer2016tutorial,rolim2019network}. This allows researchers to compare, for instance, the epistemic frames of high- and low-performing students, providing insights into how they connect ideas differently \citep{zhang2022understanding}. ENA captures the associational structure of concepts at an aggregated level \hl{and compares groups that are defined prior to the analysis} \citep{shaffer2016tutorial, gavsevic2019sens}. The SENS method \citep{gavsevic2019sens} applies both social network analysis (SNA) and ENA with ERGMs and clustering to analyze the socio-cognitive dimensions of collaborative learning processes. It provides complementary information but no unified model of heterogeneous relationships in learning processes. Building directly upon the ENA framework, Ordered Network Analysis (ONA) introduces a refinement of ENA by incorporating the sequence of codes within the defined window, thereby adding directionality to the connections and offering insight into the order in which ideas tend to arise, rather than just their reciprocal co-occurrence\hl{, while retaining ENA's shared projection space and group comparison logic} \citep{tan2022ordered}. The time window used to determine the edges in ENA and ONA defines which conceptual connections count as meaningful, so the window must be justified theoretically for the resulting network to be interpretable and accurately reflect the unit of cognitive work being analyzed. Recently, Transition Network Analysis (TNA) offers a new network-based approach focused on analyzing the sequential patterns of events in learning processes \citep{saqr2025mapping}. TNA constructs a directed, weighted network where nodes represent learning activities or states, and edges represent the probabilistic transitions from one to the next. By applying stochastic process mining approaches to sequence data, TNA identifies common pathways and temporal patterns. 

\begin{table*}[t]
\centering
\caption{\hl{Analytical capabilities of HINA compared with commonly used approaches in learning analytics}}
\label{tab:method_comparison}
\small
\begin{tabular*}{\textwidth}{@{\extracolsep{\fill}} l c c c c c c c @{}}
\hline
\textbf{Capability} & \textbf{SNA} & \textbf{ENA} & \textbf{ONA} & \textbf{TNA} & \textbf{LSA} & \textbf{K-means} & \textbf{HINA} \\
\hline
Ties between entities of different types & (\checkmark) & -- & -- & -- & -- & -- & \checkmark \\
Ties between entities of same type & \checkmark & \checkmark & \checkmark & \checkmark & \checkmark & -- & -- \\
Composite entities (e.g., code paired with target) & -- & -- & -- & -- & -- & -- & \checkmark \\
Clusters of learners & \checkmark & -- & -- & -- & -- & \checkmark & \checkmark \\
Number of clusters inferred from the data & \checkmark & -- & -- & \checkmark & -- & -- & \checkmark \\
Significance test for ties & (\checkmark) & -- & -- & \checkmark & \checkmark & -- & \checkmark \\
Individual-level measures based on interaction structures & \checkmark & -- & -- & -- & -- & -- & \checkmark \\
Comparison of entities in a shared embedding space & -- & \checkmark & \checkmark & -- & -- & -- & -- \\
Temporal order of events & -- & -- & \checkmark & \checkmark & \checkmark & -- & -- \\
\hline
\multicolumn{8}{@{}p{\textwidth}@{}}{\footnotesize \hl{\textit{Note.} SNA: social network analysis; ENA: epistemic network analysis; ONA: ordered network analysis; TNA: transition network analysis; LSA: lag sequential analysis; K-means: feature vector clustering. (\checkmark) marks a capability available through add-on methods, including two-mode networks, ERGMs, or permutation tests for SNA, which are not designed to address the same analytical questions as HINA. Details are discussed in the main text.}} \\
\end{tabular*}
\end{table*}

\hl{Table}~\ref{tab:method_comparison} \hl{summarizes the analytical capabilities of HINA compared with commonly used approaches in learning analytics. Across these approaches, no existing method is designed specifically to model and analyse heterogeneous interactions in learning processes as motivated by learning analytics needs. HINA differs most fundamentally from the non-network methods in learning analytics. Feature-based clustering such as K-means can be applied to learners' interaction counts, but these methods require the number of clusters to be fixed in advance and do not explicitly account for relational structure in interactions. Meanwhile, sequential methods such as lag sequential analysis operate on the order of coded events rather than on ties to entities. These methods answer questions about feature similarity or temporal order instead of the relational structure of heterogeneous interactions, which is HINA's focus. Below we elaborate on the need for HINA in the context of existing network-based methods. We focus on two dimensions---HINA's data representation and its analytical methods---to clarify how it differs from existing approaches. 

\textbf{\textit{Differences from existing bipartite network analyses}} For readers familiar with network analysis, a natural question is whether HINA is simply an application of existing bipartite (two-mode) network analyses, since bipartite networks can also represent interactions among different types of entities. It is not, for two reasons. First, the bipartite structural representation alone does not supply the analytical methods needed to examine that structure. A two-mode network specifies which entities are connected, but not which of those connections are statistically meaningful, how entities cluster, or how many clusters exist. Established methods for analysing such networks, including SAOMs and ERGMs, answer different questions. These methods are designed to model how ties form, characterizing the global (ERGM) and local (SAOM) structural processes that generate a network rather than summarizing its structure. Second, HINA permits a more general representation than bipartite networks by allowing for composite nodes, so that a single tie can be richer than a simple two-mode connection. In learning processes, what an action means often depends jointly on what a learner does and on whom or what it is directed toward. Seeking help from an AI agent and seeking help from a peer are not the same act. A standard bipartite network cannot capture this, because it links learners to a single type of entity at a time. It can record whom a student interacts with, or what behaviors a student performs, but not the pairing of the two. By allowing composite nodes, HINA can combine behaviors and targets into one entity, so that a single tie encodes who did what with which target at once. ``Seeking help from an AI,'' ``seeking help from a peer,'' and ``delegating to an AI'' then become distinct nodes, and a learner's interaction profile reflects both how active they were and whom they engaged along with how they distributed specific kinds of action across specific partners. Questions about human–AI collaborative learning most directly arise at this level, requiring multiple dimensions of interaction to be considered simultaneously.

\textbf{\textit{Differences from other network-based methods in learning analytics}} ENA and ONA project code co-occurrences constructed from sliding time windows into a shared low dimensional space, such that the network graph serves the purpose of visualization rather than as a structure to be analyzed directly, and comparison is made between positions in the embedding space. HINA, on the other hand, is developed for analyzing networks directly to provide node-level summative measures based on interaction structures, test whether a particular connection is stronger than a null model would predict, and identify groups of learners from their patterns of connection, rather than comparing predefined groups. ENA, ONA and HINA are complementary, addressing different analytical needs. TNA, meanwhile, treats the network as an object of analysis, like HINA, but is designed for a different purpose and to answer different questions. TNA constructs a directed, weighted network in which nodes are learning states or activities and edges are transitions between successive events, and it is designed to examine the temporal order of a process and which transitions occur often. In this sense, TNA's analytical focus is closely aligned with sequential analysis methods such as lag sequential analysis, in that both characterize how events follow one another over time. HINA's focus is fundamentally different, because rather than temporal order, it concerns the relational structure of interactions among different types of entities in learning processes. HINA connects learners to the specific tools, concepts, or people they interact with, addressing who engages with what, which ties are statistically meaningful, and how learners group by their interaction profiles.

\textbf{\textit{Novelty of the HINA framework}} In summary, HINA is novel in two connected ways. First, it provides a representation of learning interactions missing from existing methods, capturing ties among different types of entities, and through composite nodes encoding both the structure of the interactions and their relational content. Second, it provides the novel three-level analytical methods required by this heterogeneous interaction network representation, which can address questions such as how each learner participates across their interactions, which interactions are stronger than chance rather than merely observed, and how learners group. Because these methods are motivated by the questions learning analytics asks, they remain learner-centered throughout. All of these methods operate on one shared representation, so that HINA supports a systematic, multi-level analysis of learning processes within a single workflow. These methods are made available directly in the HINA Python package and web application.}


\section{Methods}

\subsection {HINA Framework}

\begin{figure*}[htbp]
    \centering
    \includegraphics[width=0.95\textwidth]{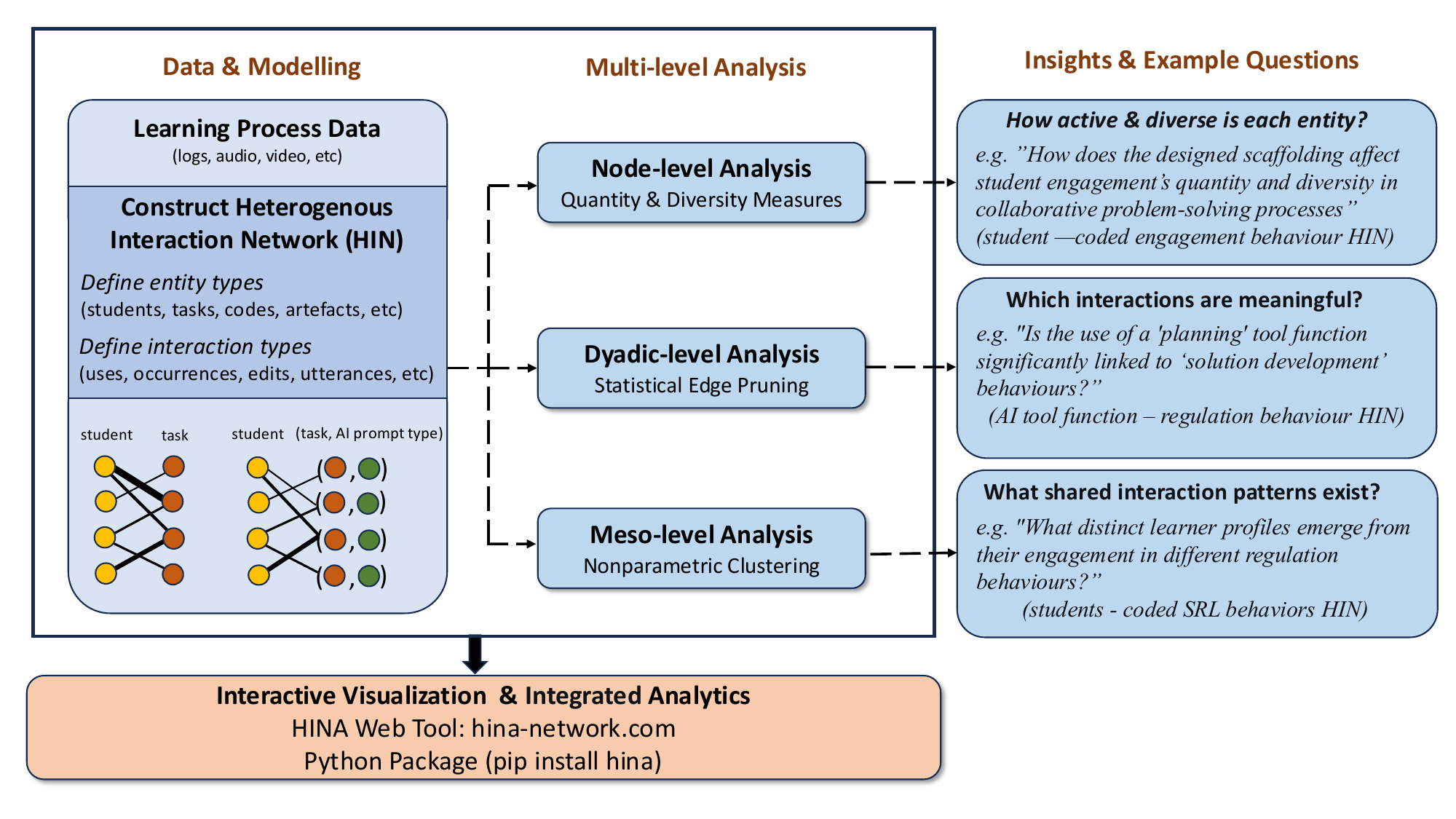}
    \caption{The HINA multi-level analytical framework}
    \label{fig:HINA_framework}
\end{figure*}

The HINA analytical framework consists of three steps: 1) Network construction and conceptual grounding; 2) Multi-level network analytics; and 3) Interactive visualization (see Figure~\ref{fig:HINA_framework}). All steps of the analysis are accessible through the HINA web tool (\href{http://hina-network.com}{\textbf{hina-network.com}}) and the HINA Python package (\texttt{pip install hina}) \citep{feng2025hina}.

\textbf{\textit{Network construction \& conceptual grounding}}: This initial step transforms raw learning process data, such as trace data collected through system logs, audio/video recordings \citep{feng2024heterogenous} into a conceptually meaningful HIN. 
\begin{enumerate}

    \item \textit{Entity Typing}: Defining the nodes $V_1$ and $V_2$, e.g., students, learning objects, coded theoretical constructs, or some combination therein for composite nodes. HINA allows for the creation of composite entities, moving beyond the limitations of simple bipartite models to capture triadic relationships (e.g., Student-(Object,Behavior)).

    \item \textit{Interaction Determination}: Using empirical observations to define meaningful edges $E$ containing the heterogeneous interactions among nodes in $V_1$ and $V_2$. Each edge indicates a direct, observable affiliation or association, e.g., a ``student performs a behavior'' or a ``student uses a tool''. 

    \item \textit{Weight Assignment}: Applying a weight function $\omega$ to edges, typically based on interaction frequency or a preconceived measure of interaction strength. The weights carry information about both the topology and intensity of heterogeneous interactions.
    
\end{enumerate}

\textbf{\textit{Multi-level network analytics}}: This step involves a set of network analytical methods for extracting results at multiple levels of granularity from the constructed HINs. \textbf{Figure \ref{fig:HINA_framework}} below presents a summary of the HINA multi-level analytical framework, including example questions for each analytical level. The design of HINA's multi-level analytical workflow is motivated by the need to move learning analytics from descriptive pattern visualization to a formal framework that supports theory-based hypothesis testing and data-driven discovery. Current approaches often culminate in visual summaries (e.g., network plots, sequence diagrams) or isolated statistical outputs, but lack a unified analytical pipeline that integrates data modeling, quantification, and statistical inference into a coherent workflow. HINA addresses this by providing the full pipeline in one framework. This allows researchers to integrate evidence from multiple analytical levels (quantitative measures, statistically significant edges, non-parametric clustering and interactive visualization) into a single account of a learning process, and to build theory from it. The multi-level analytical framework within HINA is as follows: 

\begin{enumerate}

    \item \textit{Node-level Analysis}: Calculating individual measures (the quantity and diversity of heterogeneous interactions) for each focal node in $V_1$ (typically students) to quantify the role and engagement of these entities in a learning process. This distinguishes two dimensions of engagement that the frequency-based analyses commonly employed in existing methods collapse into one.

    \item \textit{Dyadic-level Analysis}: Performing statistical testing to identify edges that are statistically significant under a specified null model. This allows for the separation of signal from noise in the underlying interaction networks, which is absent in existing network analysis methods in learning analytics. 
    
    \item \textit{Meso-level Analysis}: Applying a nonparametric clustering method to identify groups of nodes in $V_1$ that share similar heterogeneous interaction patterns or behavioral profiles. The nonparametric nature of the HINA clustering algorithm allows for inferences to be made directly from the structural regularities in the data itself, rather than requiring manual hyperparameter tuning which is common to many existing clustering methods (e.g. k-means clustering). 
    
\end{enumerate}
The multi-level analytical approach in HINA moves from analyzing individual nodes to revealing systematic structural regularities in learning process data at larger scales and validating the significance of observed interactions. HINA's multi-level analytical framework is designed as a unified system, in which the three levels together address a coherent set of research questions within a single study (as demonstrated by the case study later on). Studies can also apply any single level of analysis or any combination of these analyses to address targeted questions, either on their own or integrated with other learning analytics methods. (An example of this is in the study \citep{wong2025scaffolding}.) HINA therefore suits both application-oriented and theory-building investigations.

HINA provides an end-to-end methodology supported by an originally developed web tool (\textbf{hina-network.com}) that performs the three-level analysis and enables interactive visualization. The web tool transforms HINA's multi-level analytical outputs into an interactive visualization for discovery, interpretation, and intervention. Its customized, interactive interface allows researchers and practitioners to actively explore heterogeneous interaction patterns by enabling them to, for example, customize node and edge colors, select and highlight specific elements, drag and optimize node positions, and dynamically filter the networks. This step closes the learning analytics loop \citep{gasevic2019we, clow2012learning}, turning analytical outputs into a basis for pedagogical decisions and intervention. The HINA web tool functions as a research analytics tool for conducting the three-level analysis and visualization, and can also be used as a learning analytics dashboard to present to students and teachers for pedagogical intervention. 

\subsection {Methodological Details}\label{sec:methods_details}
\subsubsection {Individual-level analysis}\label{sec:individual}

HINA provides two individual-level measures, quantity and diversity of interactions, for analyzing heterogeneous interaction networks. The quantity measure reflects the overall intensity or frequency of a focal entity's heterogeneous interactions, and the diversity measure reflects the variety of those interactions across the available target entities. For instance, in a student-artefact HIN, a student's quantity would reflect their total level of engagement with learning artefacts (e.g., number of interactions), while their diversity would indicate whether that engagement was spread evenly across many artefacts or concentrated on just a few \citep{Feng2025JLA}. 

The quantity measure $Q_i$ for node $i\in V_1$ quantifies $i$'s overall interaction propensity, calculated as the normalized sum of interaction weights between node $i$ and all nodes in $V_2$. Formally, for total network weight $W = \sum_{i\in V_1}\sum_{j\in V_2}w_{ij}$, the quantity measure is defined as
\begin{align}
\label{eq:quantity}
~~~~~~~~~~~~~~~~~~~~~~~~~~~~~~~~~~~~~~~~~~~~~~~~~~~~~~~~~~~~~~~~~~~~~~~~~~~~~~Q_i = \frac{1}{W}\sum_{j\in V_2}w_{ij}.
\end{align}
This measure can be further normalized within specific subgroups $g\subseteq V_1$ using $W_g = \sum_{i\in g}\sum_{j\in V_2}w_{ij}$ to yield
\begin{align}
~~~~~~~~~~~~~~~~~~~~~~~~~~~~~~~~~~~~~~~~~~~~~~~~~~~~~~~~~~~~~~~~~~Q^{(g)}_i = \frac{1}{W_g}\sum_{j\in V_2}w_{ij} = \frac{W}{W_g}Q_i.
\label{eq:quantity_group}
\end{align}
While quantity captures engagement volume, it does not characterize how heterogeneous interactions are distributed across different nodes $j\in V_2$. The diversity measure $D_i$ addresses this by quantifying the heterogeneity of interaction patterns using a normalized variant of the Shannon entropy \citep{shannon1948mathematical}, given by
\begin{align}
\label{eq:diversity}
~~~~~~~~~~~~~~~~~~~~~~~~~~~~~~~~~~~~~~~~~~~~~~~~~~~~~~~~~~~~D_i = -\frac{1}{\log N_2}\sum_{j\in V_2}\left(\frac{w_{ij}}{s_i}\right)\log \left(\frac{w_{ij}}{s_i}\right),
\end{align}
where $N_2 = \vert V_2\vert $ and $s_i = WQ_i = \sum_{j\in V_2}w_{ij}$ is node $i$'s total (unnormalized) interaction strength. We use the shorthand $\log \equiv \log_2$, although it does not actually make a difference which base is used due to the normalization. The diversity measure ranges from $D_i=0$ (all of $i$'s interactions concentrate on a single node $j\in V_2$) to $D_i=1$ ($i$'s interactions are uniformly distributed across all available nodes in $V_2$)\hl{, with the convention $0\log 0=0$ and $N_2\geq 2$. Note that $D_i$ measures how evenly interactions are spread over the targets and is symmetric in the targets, so when $N_2=2$ two students who split their interactions in ratios $3{:}1$ and $1{:}3$ have the same $D_i$ (i.e., which target dominates must be read from the counts themselves)}. 

The quantity and diversity   measures capture fundamental, distinct dimensions of an entity's heterogeneous interactions by reflecting interaction intensity (quantity) and interaction breadth (diversity), respectively. High quantity does not imply high diversity, and vice versa. The two measures therefore carry different information than a simple interaction count. For example, in a student-artefact HIN, a student could exhibit high-quantity, low-diversity engagement (intense, focused use of one artefact) or low-quantity, high-diversity engagement (exploratory, distributed use of many artefacts) \citep{Feng2025JLA}. By analyzing both measures, researchers can construct detailed profiles of how individuals (or resources, tools, or behaviors) function within a learning environment. These quantitative profiles, in turn, enable hypothesis testing about how contextual factors shape interaction patterns and provide a basis for theorizing emerging roles and examining the effects of interventions on participation quantity and diversity \citep{Feng2025JLA, wong2025scaffolding}.

\subsubsection {Dyadic-level analysis}\label{sec:dyadic}
The dyadic-level analysis in HINA identifies statistically significant heterogeneous interactions within the constructed HINs. A study by \citet{serrano2009extracting} introduces a multiscale ``disparity filter'' that extracts a network's essential backbone by using a local statistical significance test to preserve heterogeneously distributed edges across all scales, effectively maintaining the system's topological and hierarchical features where global thresholding methods fail. Building on this backboning principle, the HINA framework proposes dyadic-level analysis methods for pruning heterogeneous interaction networks based on various null models. \textbf {Dyadic-level analysis moves beyond describing what interactions exist to evaluating which interactions are meaningfully structured by the learning process}. At this level, we test whether an observed connection between two specific entities (e.g., a student and a particular tool) occurs more often than expected by chance, which would suggest a deliberate or pedagogically meaningful relationship, or whether it is a random or incidental co-occurrence. \textbf{The significance of an interaction is determined by comparing its observed frequency (edge weight) against a null model that simulates a plausible random baseline for connectivity within the HIN}. This process statistically `prunes' the network, filtering out noise to leave the backbone of heterogeneous interactions that characterize the learning system. 

Given an edge $(i,j)$ in an HIN, one is often interested in determining if this edge represents a statistically meaningful interaction among $i$ and $j$ during the learning process, or instead a spurious or transient passing interaction. Whether or not $(i,j)$ is statistically significant depends on its weight $w_{ij}$ and the null model one chooses for the connectivity structure of the HIN. HINA offers methods to prune HINs under multiple different null models for maximum flexibility to serve different research questions. 

A natural null model for comparison is one in which the edges and total weight $W=\sum_{ij}w_{ij}$ are placed completely at random between the nodes in the two node sets. In HINs, each edge weight $w_{ij}\geq 1$ is a positive integer, so a single edge of weight $w_{ij}$ can be viewed as $w_{ij}$ edges of weight one. We can then consider a null model in which $W$ edges of weight one are distributed uniformly at random with replacement among the $N_1N_2$ possible node pairs in the HIN. Defining
\begin{align}
~~~~~~~~~~~~~~~~~~~~~~~~~~~~~~~~~~~~~~~~~~~~~~~~~~~~~~~~~~~~~~~~~~~~~~f_{n,\rho}(k) = {n\choose k}\rho^{k}(1-\rho)^{n-k}
\end{align}
as the Binomial probability mass function (PMF) with $n$ trials and success probability $\rho$ (with $f_{n,\rho}(k)=0$ for $k<0$ or $k>n$), we have that the probability mass function for the weight $w_{ij}$ is $f_{W,\rho_0}(w)$ under this null model, with
\begin{align}
~~~~~~~~~~~~~~~~~~~~~~~~~~~~~~~~~~~~~~~~~~~~~~~~~~~~~~~~~~~~~~~~~~~~~~~~~~~~~~~~~~~~\rho_0=\frac{1}{N_1N_2} 
\end{align}
the probability of each uniformly random edge placement landing on any particular node pair $(i,j)$.

Given this null model, we can define a statistically significant edge $(i,j)$ as one for which the probability of obtaining a weight \emph{at least as high} as $w_{ij}$ under the null model is less than some significance level $\alpha\in (0,1)$. This translates to keeping $w_{ij}$ in the HIN as long as
\begin{align}
~~~~~~~~~~~~~~~~~~~~~~~~~~~~~~~~~~~~~~~~~~~~~~~~~~~~~~~~~~~~~~~~~~~~~~~~~~~~~~w_{ij}> q_{W,\rho_0}(1-\alpha),    
\end{align}
where $q_{n,\rho}$ is the quantile function associated with the Binomial PMF $f_{n,\rho}$. The default pruning option HINA offers (\text{fix\_deg} = \text{``None''}) uses this weight threshold, pruning edges with $w_{ij} \leq q_{W,\rho_0}(1-\alpha)$. This can easily be computed using standard statistical software packages. Here $N_1$, $N_2$ and $W$ always refer to the network being pruned: when the dyadic analysis is applied to a projected network, such as the code--target networks of individual clusters in Section~\ref{sec:results}, they are the numbers of code nodes, target nodes and interactions in that projected network.

The default null model above assumes that $w_{ij}$ follows a Binomial distribution with $\rho_0=(N_1N_2)^{-1}$ a constant for all node pairs $(i,j)$. In other words, all edge weights are distributed uniformly at random across all valid node pairs. However, in practice there is often substantial heterogeneity in the node strengths $s_i=\sum_{j}w_{ij}$ and $d_j=\sum_{i}w_{ij}$, such that particular pairs $(i,j)$ are more likely to have high edge weights $w_{ij}$ due to $i$ and/or $j$ having a high strength. HINA offers the optional argument $\text{``fix\_deg''}$ to modify the above Binomial null model by adjusting $\rho$ based on the strengths in either of the node sets $V_1$ or $V_2$. For example, if the first node set of size $N_1$ contains the `student' nodes, and we want to condition our null model on the overall level of participation of each student, we can fix the student degrees in the null model by setting $\text{`fix\_deg'}=\text{`student'}$. The underlying null model in this case is then still Binomial, but now with the number of trials set to $s_i=\sum_{j}w_{ij}$ and the success probability equal to $\rho_{2}=N_2^{-1}$ for each edge connected to student $i$. Similarly, we can fix the node degrees in $V_2$, which sets the number of trials to $d_j=\sum_{i}w_{ij}$ and the success probability to $\rho_{1}=N_1^{-1}$ for each edge connected to target node $j$. As before, edges with weights at or below the threshold $q_{n,\rho}(1-\alpha)$, with $n,\rho$ modified appropriately for the null model of interest, are pruned to highlight statistically significant edges.

The interpretation of pruned networks depends fundamentally on the null model construction. When strengths are not fixed for either node set, the pruned network reveals heterogeneous interactions that are significant relative to the interaction frequency observed across the entire HIN. In contrast, when strengths are fixed for a node set $V_a$, the pruned network reveals heterogeneous interactions that are significant relative to the interaction frequency observed at each node $i\in V_a$, controlling for their overall activity levels. The choice between these approaches should be driven by the research question and the weight heterogeneity in the HIN. For example, if a researcher observes that the quantity of engagement $Q_i$ across students $i\in V_1$ is highly heterogeneous and hypothesizes that this heterogeneity can be attributed to unobserved student attributes (e.g. intrinsic motivation for the learning task), then the null model in which the strength $s_i=WQ_i$ of each student node is fixed is most appropriate as it can identify interactions that are significant for each student given their overall level of engagement. In this case, one can set $\text{`fix\_deg'}=\text{`student'}$ in the HINA tool. On the other hand, if the engagement strength $s_i$ does not exhibit substantial variability across nodes and/or both node sets are of equal focus in the analysis (often the case when neither node set contains students), then the default null model in which no strengths are fixed is most appropriate as it treats the node sets symmetrically. 

For an alternative visualization of the pruned networks, the HINA webtool also offers the option to compute a one-mode projection of the dyadic graph onto either node set, assigning the edge weight for a node pair as its cosine similarity \citep{newman2018networks}. If the dyadic network has been pruned ahead of this one-mode projection, the resulting network reflects the more stringent criterion for shared interactions. This allows for more compact visualization when the projected-out node set is large, and also helps to highlight mesoscale regularities.

\subsubsection {Meso-level analysis}\label{sec:meso}

The meso-level clustering analysis addresses a specific, common analytical need in learning analytics: \textbf{to group entities from a single focal set (e.g., students) based on the patterns of their heterogeneous interactions with other entity types (e.g., tools, behaviours, resources).} To identify clusters of nodes with similar interaction patterns in HINs, HINA employs a new nonparametric community detection method which acts on only a single node set to identify its internal structural regularities.  

Standard clustering algorithms (e.g., K-means) operate on feature vectors, treating heterogeneous interactions as independent attributes rather than as an interconnected relational system. Although a multitude of community detection methods exist for networks \citep{FORTUNATO201075}, they are predominantly designed for homogeneous networks in which nodes are of a single type. While clustering methods also exist for heterogeneous information networks \citep{wei2021multiple} and bipartite/multipartite networks \citep{yen2020community} consisting of multiple distinct node types, they do not explicitly extract clusters based on structural regularities shared among nodes only in one node set, instead focusing on both node sets simultaneously.

Existing approaches do not readily address the specific problem we target here:  \textbf{nonparametric clustering of one node set in a bipartite setting based on similarity in interaction profiles}. Within the HINA framework, we cluster nodes within only the \emph{focal node set} (the node set of interest for clustering, typically a student node set), rather than simultaneously grouping nodes from both sets. By doing so, the interaction patterns across the distinct, ungrouped nodes in the second set defines the membership of the focal node set. This design is essential for many learning analytics applications, such as clustering students according to their fine-grained engagement with individual coded behaviors or specific learning resources\citep{heinimaki2021student, li2024individuals}, where preserving the identity and independence of the second set of nodes is important for meaningful interpretation.

Existing stochastic block models and other community detection methods focus on connectivity at the group-level such that an edge $(p,q)$ and an edge $(p,q')$ are treated identically if $q$ and $q'$ are both part of the same cluster $Q$. This means that blockmodels would fail to explicitly extract clusters based on shared interaction structure at the node-level between two nodes $p,p'$ in a group $P$ since the existence of the edges $(p,q)$ and $(p',q')$ (with $p$ and $p'$ not sharing a target node) is identical to the existence of the edges $(p,q)$ and $(p',q)$ (with $p$ and $p'$ sharing a target node) from the blockmodel's point of view, as it just sees two ties from group $P$ to group $Q$ in both cases. This motivates the construction of a new principled method for clustering nodes in one set based on their shared regularities with individual nodes in a second set. 

In HINA, we propose a new clustering method, grounded in the Minimum Description Length (MDL) principle from information theory \citep{rissanen1978modeling}. The MDL principle states that the best model for a dataset is the one that allows for its shortest description in terms of bits of information. The MDL principle has been successfully applied in a variety of network partitioning contexts, including (unipartite) community detection \citep{peixoto2017nonparametric}, core-periphery and hub partitioning \citep{gallagher2021clarified,kirkley2024identifying}, spatial regionalization \citep{kirkley2022spatial,morel2024bayesian}, and rank clustering \citep{morel2025estimation}. In HINA framework, we employ the MDL principle to address the methodological gap by enabling focused clustering of a target node set while fully accounting for the heterogeneity of its connections to other entity types, which is what many learning analytics scenarios require. For example, if we aim to identify students with similar engagement patterns during collaborative problem solving (CPS) processes \citep{Feng2025JLA}, we only need to cluster student nodes, using the nodes in the second node set purely as a means to identify similar students. Without loss of generality, we can let $V_1$ be the set of nodes we want to cluster. 

The community detection method finds a community partition of the nodes in $V_1$, automatically selecting for the optimal number of clusters by employing the MDL principle for regularization. A community partition $\bm{b}$ corresponds here to a vector of $N_1=\vert V_1\vert$ node labels, with $b_i$ the community label of node $i\in V_1$. We will denote with $B$ the number of unique labels (groups) in the partition $\bm{b}$, and $n_r$ the number of nodes in group $r\in \{1,...,B\}$ according to the partition $\bm{b}$. We will also let $w^{(cn)}_{rj}$ be the total weight of edges running from community $r$ to node $j\in V_2$. 

The MDL objective we develop considers the total number of bits of information $\mathcal{L}(G,\bm{b})$ required to transmit the HIN $G$ by using the node partition $\bm{b}$ to coarse-grain the structure of the HIN. $\mathcal{L}(G,\bm{b})$ is called the \emph{description length} of the data $G$ under the partition $\bm{b}$. The number of bits $\mathcal{L}(G,\bm{b})$ required to transmit $G$ using a partition $\bm{b}$ of the nodes in $V_1$ can be computed using fixed-length codes \citep{cover1999elements}, commonly employed in MDL applications within network science \citep{peixoto2017nonparametric,kirkley2022spatial,kirkley2023compressing,kirkley2024paninipy}. 

The derivation of $\mathcal{L}(G,\bm{b})$ with fixed-length codes is given in Appendix~\ref{app:derivation}; the resulting description length is

\begin{align}
\label{eq:MDLobj}
~~~~~~\mathcal{L}(G,\bm{b}) &= \log N_1 + \log {N_1-1\choose B-1}
+\log {N_1\choose n_1,...,n_B} \\
&+ \log {BN_2+W-1\choose W}
+\sum_{r=1}^{B}\sum_{j=1}^{N_2}\log {n_r+w^{(cn)}_{rj}-1\choose w^{(cn)}_{rj}}\nonumber.
\end{align}
We can now minimize this clustering objective over partitions $\bm{b}$ to determine the optimal partition $\bm{b}$ of the nodes in the first node set according to the MDL principle.

Eq.~\ref{eq:MDLobj} scores a partition $\bm{b}$ of the nodes in $V_1$ according to how well it compresses the information in $G$ by exploiting structural regularities in the edges incident on each community in $\bm{b}$. If each community $r=1,...,B$ has very homogeneous target nodes $j$ in the second node set $V_2$ (i.e., nodes in group $r$ allocate most of their interactions to a small subset of nodes in $V_2$), then the description length will decrease. On the other hand, if the communities do not capture regularities in the connections (i.e. the connectivity patterns among nodes in group $r$ are highly diverse), the description length will increase. While it is often hard to capture such regularities with few groups $B$, the MDL objective penalizes us for having too many groups since it is more complex to describe the corresponding partitions $\bm{b}$. Thus, under the objective in Eq.~\ref{eq:MDLobj}, partitions $\bm{b}$ should be made more complex, i.e. contain more groups $B$, only as long as these additional groups capture enough additional structural regularity among the nodes in $V_1$ to justify the increased complexity of $\bm{b}$ from an information theoretic point of view. Minimizing Eq.~\ref{eq:MDLobj} therefore returns a partition $\bm{b}$ that parsimoniously groups the nodes in $V_1$ into clusters whose interaction patterns with nodes in $V_2$ are similar enough to shorten the description of the network. No user input is needed to determine the number or composition of the node clusters obtained by this mesoscale analysis, so the clusters adapt to the structural variability in the data.

To optimize Eq.~\ref{eq:MDLobj} HINA employs an efficient agglomerative greedy merging scheme in which all nodes start off in their own cluster, and we iteratively merge the clusters of nodes that provide the greatest reduction in the description length (or, smallest increase, if no merge decreases $\mathcal{L}$). We store the description length at all values of $B$ until the algorithm terminates at $B=1$, then check these past solutions to find the optimal value of $B$ and its corresponding partition $\bm{b}$. Such greedy approximate optimization schemes have been shown to have good performance relative to exact enumeration for previous MDL applications in the network context \citep{kirkley2024inference,kirkley2025fast}. The scheme is deterministic given the input, evaluates $O(N_1^2)$ candidate merges at the first step and $O(N_1)$ at each subsequent step, each in $O(N_2)$ time, and runs in well under a second for the case-study network. All analyses reported in this paper were produced with version 0.7.5 of the HINA Python package. Benchmarking against ground truth labels is not a good way to validate unsupervised clustering methods in this context, because empirical metadata need not align with network structure \citep{peel2017ground} and synthetic benchmarks can be trivially optimized by their own generative models \citep{peixoto2023implicit}. Instead, we establish construct validity by arguing that our method captures the intended similarity in Eq.~\ref{eq:MDLobj}, \hl{deferring a numerical comparison to Appendix}~\ref{app:comparison}\hl{ as a secondary evaluation criterion}.

After minimizing the description length, we can extract a \emph{compression ratio} telling us how well the clusters reduce the description length of the HIN beyond a trivial partition (i.e. when all nodes are in the same cluster). We define the compression ratio here as
\begin{align}
~~~~~~~~~~~~~~~~~~~~~~~~~~~~~~~~~~~~~~~~~~~~~~~~~~~~~~~~~~~~~~~~~~~~~~~~~~~~\eta = \frac{\mathcal{L}(G,\bm{b}_{\text{MDL}})}{\mathcal{L}(G,\bm{b}_{0})},
\end{align}
where $\bm{b}_{\text{MDL}}=\argmin_{\bm{b}}\{\mathcal{L}(G,\bm{b})\}$ is the minimum description length solution for the node partition (as approximated by the proposed merging algorithm), and $\bm{b}_0$ is the trivial partition in which all nodes belong to the same cluster. \hl{The same ratio $\eta(\bm{b})=\mathcal{L}(G,\bm{b})/\mathcal{L}(G,\bm{b}_0)$ can be computed for any partition $\bm{b}$, for example one produced by another method, and may then exceed one.} \hl{Since the merging algorithm always retains the trivial partition as a candidate, $\eta\in(0,1]$, with $\eta=1$ meaning that no partition compresses the network and smaller values meaning greater compression. The absolute value of $\eta$ is not directly comparable across datasets, since the description length is dominated by the cost of transmitting the edge weights, which grows with $N_1$, $N_2$ and $W$, so the same underlying regularity produces a value much closer to one in a network of tens of nodes than in a network of thousands of nodes} \citep{morel2024bayesian}\hl{. Appendix}~\ref{app:comparison} \hl{reports a permutation test of the value of $\eta$ obtained for the case study, as a possible way of checking that a given value reflects more than the heterogeneity of node strengths.}

\subsubsection {\hlsec{Comparison of HINA meso-level analysis with existing community detection methods in network analysis}}\label{sec:louvain}

\hl{Community detection is a critical technique in network analysis for clustering nodes based on structural similarities. This subsection clarifies how the HINA meso-level clustering in Section}~\ref{sec:meso} \hl{differs from the existing community detection methods}. \hl{Given the analytical need in HINA is to cluster focal nodes based on their interactions with different types of entities, there are two streams of comparable existing clustering methods in network analysis.} \hl{The first stream of methods manages clustering directly on bipartite structure graphs, by clustering both node sets at once, either by maximizing a bipartite modularity} \citep{guimera2007module} \hl{or by fitting a stochastic block model} \citep{peixoto2017nonparametric}\hl{. As discussed above, Such methods cluster nodes on both sides of a bipartite structure, for instance, grouping the learners and the artefacts they act on simultaneously. This shows which cluster of students interacts with which cluster of artefacts, but it does not cluster students based on their interaction patterns with individual artefacts. As a result, two learners who direct their interactions at different artefacts within the same artefact cluster become indistinguishable. HINA, by contrast, clusters only the focal node set based on its heterogeneous interaction patterns while leaving the target node set unpartitioned. Thus, in the case of students interacting with artefacts, HINA can cluster learners by the specific artefacts they worked on rather than only identifying which broad groups of students and artefacts are associated. 

The second stream consists of clustering methods designed for networks with a single node type. Applying them to a HIN therefore requires first projecting the constructed HIN onto the focal node set, weighting each pair of learners by the interactions they share, and then running a standard one-mode method on the projected network (most commonly modularity maximization with the Louvain algorithm} \citep{blondel2008fast}\hl{). To make a direct comparison with this stream of methods, we ran the Louvain algorithm on the case study data for four common projection weightings and three values of the resolution parameter (Appendix}~\ref{app:comparison}\hl{). The number of communities returned by Louvain depends on the resolution parameter, with one, two, or between six and thirteen communities returned on the same network at the three settings, whereas the number of clusters in HINA is inferred from the data itself. The projection also loses the information that the clustering is meant to use, since 84\% of learner pairs share at least one (code, target) entity and the projected network is therefore almost complete, with no record of which entities each pair shares. When all partitions are scored with the description length of Eq.}~\ref{eq:MDLobj}\hl{, none of the Louvain partitions compresses the HIN by more than 1/10 of a bit, while the HINA partition provides significantly greater compression than expected under a randomized null model} (Appendix~\ref{app:comparison}).

\section {Case Study: Applying HINA to Model Human-AI Interactions in Small-Group Collaborative Learning}\label{sec:casestudy}

In this section, we present a case study to demonstrate the analytical capabilities of the HINA framework by modeling human-AI interactions in a small-group collaborative learning environment in which each group of students interacts with an AI chatbot. We employ HINA's core three-level analytical workflow (individual, dyadic, and mesoscale analysis) to reveal the underlying interaction patterns. The analysis below focuses on the contextualization and interpretation of the network measures within the case study, rather than their detailed methodological formulation, which is provided in Section~\ref{sec:methods_details}. By applying HINA to these data, we aim to provide a methodology that lets future research connect these new measures with group or individual-level metadata for advancing human-AI collaboration theory development. 

The case study we present addresses the following research questions:

\textbf{RQ1}: How do students vary in their interaction profiles within the AI-assisted collaborative environment, as measured by the quantity and diversity of their engagement with peers versus the AI agent? 

\textbf{RQ2}: What distinct engagement patterns emerge when clustering students based on their profiles of student–AI and student–peer interactions across cognitive, metacognitive, socio-emotional, and coordinative dimensions?

\textbf{RQ3}: What specific configurations of student–AI and student–peer interactions, across the cognitive, metacognitive, socio-emotional, and coordinative dimensions, characterize and differentiate each emergent engagement pattern?

\subsection{Study Context and Participants}

This case study analyzes data from a two-week collaborative learning task in an undergraduate project management course on information management in Hong Kong. Twenty-seven students participated, forming six groups of 3-5 members. Their objective was to collaboratively analyze a complex problem and produce a written report. Student groups were required to complete this task as part of their group project in the course. The collaboration was facilitated by an online messaging tool integrated with a Generative AI (GAI) chatbot, which groups could use autonomously. The chatbot was powered by GPT-4 and configured with course-specific knowledge to respond to queries, summarize discussions, and provide feedback. All interaction data was collected through the online messaging tool. 

\subsection{Data Collection and Preprocessing}

The dataset for this analysis comprises the complete log of interaction events between students and the GAI chatbot, as well as among students themselves, recorded over the two-week period. Each row of the process data represented a single message instance, with columns including `Group info', `user\_id' (the individual sending the message), `utterance' (the raw text content of the message), `engagement codes', and `interaction target' (Peer or AI). We developed a coding framework, grounded in the CoMPAS model for individual participation in collaborative learning \citep{feng2025compas}, to analyze the collaborative learning processes. The CoMPAS framework includes both individual solidarity participation (including passive participation and active contributions that do not involve interacting with others) and interactive participation (which involves engaging with other group members). As this study focuses on students' interactions with AI and their group peers, we use the four dimensions of interactive participation from the CoMPAS framework to analyze group dynamics in AI-assisted collaborative learning. This framework categorizes group interactions into four distinct dimensions in collaborative learning: Cognitive -- task-focused knowledge construction; Metacognitive -- process-focused planning and monitoring; Socio-Emotional -- focused rapport building; and Coordinative -- task and resource management. In this study, each dimension was operationalized into specific, well-defined codes with clear examples, as detailed in Table \ref{tab:coding}. Two trained annotators applied this coding framework to code the complete two-week message log, achieving high inter-rater reliability (Cohen's $\kappa$ = 0.92). This preprocessing yields the coded interactions between students, their peers and the AI chatbot that the following analyses use.

\begin{table*}[t]
\centering
\caption{The Coding Framework for Group Interactions in Collaborative Learning}
\label{tab:coding}
\small
\begin{tabular}{p{0.11\linewidth} p{0.16\linewidth} p{0.35\linewidth} p{0.21\linewidth}}
\hline
\textbf{Dimension} & \textbf{Code} & \textbf{Definition} & \textbf{Example} \\
\hline
\multirow{5}{=}{\textbf{Cognitive}} & Concept/Solution Questions & Asking for factual information, clarification, or explanation related to the task content. & "What does this term mean?" \\
\cline{2-4}
 & Answer/Explanation & Providing factual information, clarification, or explanation in response to a question. & "It means the process is cyclical." \\
\cline{2-4}
 & New Idea/Suggestion & Introducing a new concept, solution, or approach to the task. & "What if we used a graph to show this?" \\
\cline{2-4}
 & Critical Assessment & Assessing, challenging, or building upon an idea proposed by others. & "The data on page 5 contradicts that." \\
\hline
\multirow{3}{=}{\textbf{Metacognitive}} & Planning & Discussing goals, strategies, or a roadmap for completing the task. & "Let's brainstorm first, then outline." \\
\cline{2-4}
 & Monitoring & Checking progress, identifying confusion, or recognizing a problem. & "Are we on track?" "I think we're getting off-topic." \\
\cline{2-4}
 & Evaluation & Looking back on what was learned or evaluating the group's process. & "This method isn't working efficiently." \\
\hline
\multirow{4}{=}{\textbf{Socio-Emotional}} & Encouragement/ Greetings & Providing greetings, positive feedback, approval, or motivation to peers. & "hi hi", "Great idea, Joy!" \\
\cline{2-4}
 & Appreciation & Expressing gratitude for help or contributions. & "Thanks for clarifying that." \\
\cline{2-4}
 & Empathy & Acknowledging or sharing the feelings of others. & "I also found that part frustrating." \\
\cline{2-4}
 & Agreement/Alignment & Brief, affirmative responses that signal consensus, validation, or confirmation of others' comments & "Yes", "Agree" \\
\cline{2-4}
\hline
\multirow{4}{=}{\textbf{Coordinative}} & Task Clarification & Understanding instructions, scope, and formal constraints. & "How many words is the report?" \\
\cline{2-4}
 & Task Assignment & Delegating or accepting specific responsibilities. & "I'll write the introduction." \\
\cline{2-4}
 & Time Management & Setting or reminding others of time constraints. & "Let's have a draft by Wednesday." \\
\cline{2-4}
 & Resource Coordination & Discussing the use of tools, documents, or materials. & "I created a shared folder." \\
\hline
\end{tabular}
\end{table*}

\subsection{Data Analysis Using HINA}

To address the research questions, we constructed two distinct HINs from the coded interaction data. The analysis was then conducted using the HINA framework introduced in Section~\ref{sec:methods_details}. 

To model the student-peer and student-AI interaction structures and content, we created two different HINs in this study. The first, a \textbf{Student-Target HIN}, models the choice of interaction targets for the students during the collaborative learning process. Formally, we can denote this HIN as $G^{(st)} = (V^{(st)}_1, V^{(st)}_2, E^{(st)}, \omega)$, with $V^{(st)}_1$ the set of student nodes, $V^{(st)}_2$ the set of target nodes (the interaction targets for students, of two types, ``Peer'' and ``AI''), and $E^{(st)}\subseteq V^{(st)}_1\times V^{(st)}_2$ the observed heterogeneous interactions of the form (student, peer) and (student, AI). The weight $\omega((i,j))$ here is given by the number of times the student $i$ and target $j$ interacted over the study time window. This network is the basis for analyzing individual interaction strategies (RQ1).

The second HIN, a \textbf{Student-(Content, Target) HIN}, models the full complexity of interactions considering both the interaction structure and content. Formally, we can denote this HIN as $G^{(sct)} = (V^{(sct)}_1, V^{(sct)}_2, E^{(sct)}, \omega)$, with $V^{(sct)}_1$ the set of student nodes and $V^{(sct)}_2$ the set of (content,target) nodes, each of which includes a content code and an interaction target (i.e. another student or the AI). The edge set $E^{(sct)}\subseteq V^{(sct)}_1\times V^{(sct)}_2$ in this HIN contains the observed heterogeneous interactions of the form (student, content code + student) and (student, content code + AI). This time, each node $j\in V^{(sct)}_2$ is a composite entity $j=(j_c,j_t)$ consisting of a content code and a target. Thus, the edge $(i,j)$ indicates that the student $i$ interacted with the target $j_t$ using content coded as $j_c$, and the weight $\omega((i,j))$ is the number of times this happened over the study time window. The content codes comprise the 15 coded interaction behaviors along the cognitive, metacognitive, socio-emotional, and coordinative dimensions illustrated in Table~\ref{tab:coding}, of which 14 occur in Cluster 1 and 13 in Cluster 2 (Table~\ref{tab:sigedges}). This nested bipartite structure lets the analysis recover engagement patterns that depend on both interaction frequency and content (RQ2 and RQ3).

To address RQ1, which focuses on individual interaction strategies, we analyzed the student-target HIN using HINA's individual-level analysis as introduced in Section~\ref{sec:individual}. For each student node, we computed the quantity and diversity (Eq.s~\ref{eq:quantity}~and~\ref{eq:diversity}) of their connections to the AI chatbot and student partners. Within this context, \textbf{the quantity refers to the raw volume of a student's engagement, operationalized as their total number of interactions with either AI or peers}. This metric indicates the overall level of a student's activity within the collaborative environment. Meanwhile, diversity captures the distribution of a student's interactions across the available targets. This metric reveals \hl{how evenly a student's interactions were split between
peers and the AI, with 1 indicating a perfectly even split and 0 indicating a
focus on peers or the AI alone.}

To address RQ2, which seeks to identify emergent engagement patterns, we performed a mesoscale analysis on the Student-(Content, Target) HIN using HINA's mesoscale clustering as introduced in Section~\ref{sec:meso}. This allowed us to cluster students who share statistically similar configurations of student-(code,target) interactions, while automatically inferring the optimal number and composition of the clusters from the data in a nonparametric manner. The clustering results identify distinct engagement patterns: students within the same cluster have similar profiles of student–AI and student–peer interactions. Students within each cluster share common interaction frequencies with the different interaction codes (cognitive, metacognitive, socio-emotional, coordinative) and targets (AI, peers), and the interaction patterns within each cluster are distinct from those within other clusters.

To address RQ3, which aims to characterize the clusters identified in RQ2, we utilized the projected code--target networks generated by HINA's dyadic-level analysis as introduced in Section~\ref{sec:dyadic}. For each student cluster, HINA produces a simplified network showing the (content code, target) interactions aggregated over all students in a cluster. Formally, given a cluster of students $g\subset V^{(sct)}_1$, the weight $w_{ct}$ of an interaction among content code $c$ and target $t$ in the projected network for cluster $g$ is given by $w_{ct}=\sum_{i\in g}\omega((i,(c,t)))$. In words, the weight of the edge among the content code $c$ and the target $t$ in cluster $g$'s projected network is the total number of times students in $g$ interacted with target $t$ using content code $c$ during the study time window.

We then applied HINA's dyadic-level analysis to these cluster-level projected networks to identify which specific code-target edges (e.g., Concept/Solution Questions--AI, Task Assignment--Peer) were statistically significant relative to a null model of random interaction. For this analysis we did not fix the degree of either node set, so as to treat the two node types symmetrically. These significant edges support a functional interpretation of each cluster rather than a description of it. For example, a significant Concept/Solution Questions--AI association indicates that students in the current cluster systematically directed their task-related inquiries towards the AI chatbot. By analyzing these significant edges for each cluster, this approach gives an interpretable account of the engagement patterns that define the cluster.

\subsection{Results}\label{sec:results}

\subsubsection{RQ1: Individual Interaction Profiles}

\begin{figure}[htbp]
    \centering
    \includegraphics[width=0.9\columnwidth]{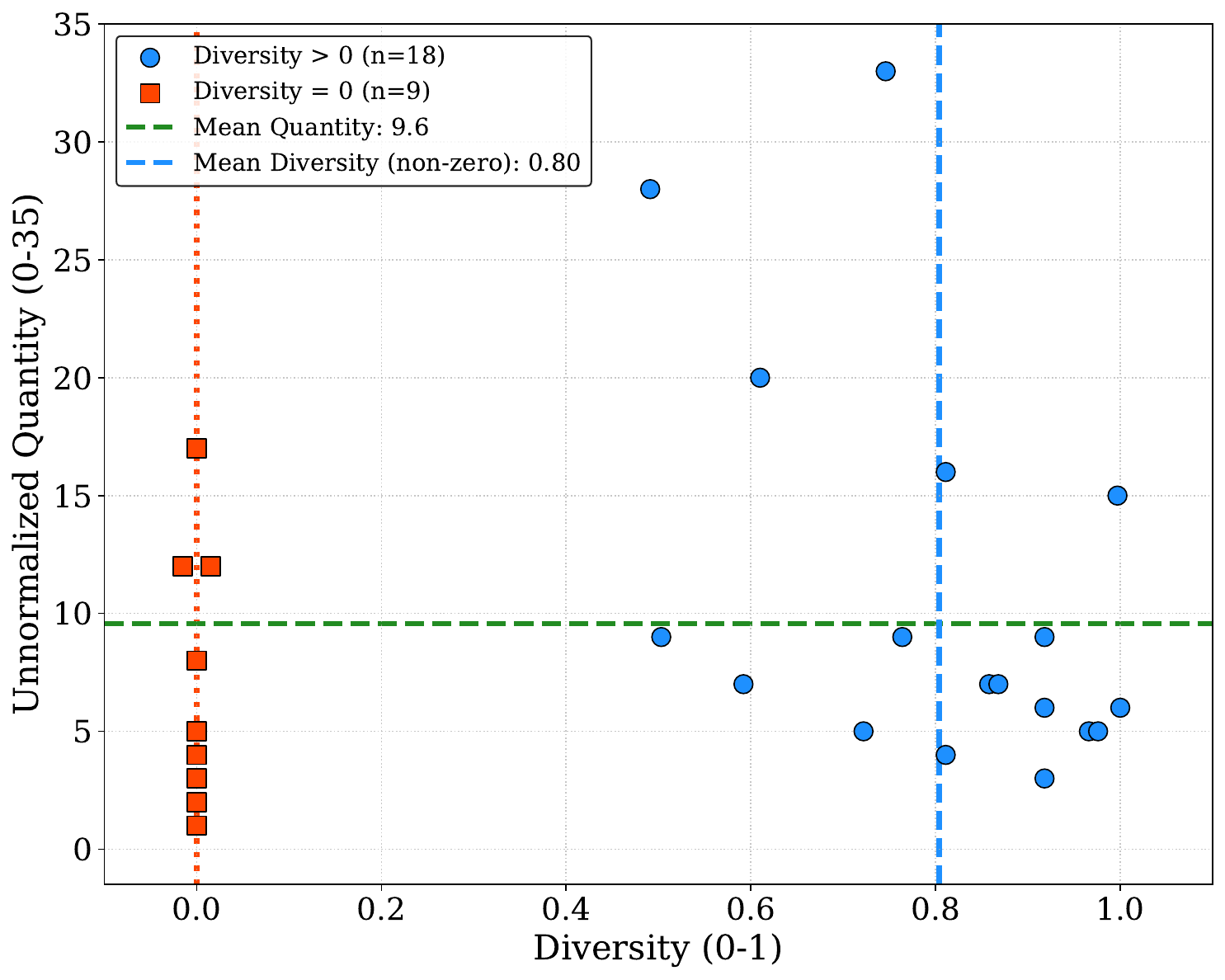}
    \caption{Individual interaction profiles of students within the student-target HIN, showing the (unnormalized) quantity $s_i=WQ_i$ (Eq.~\ref{eq:quantity}) and diversity $D_i$ (Eq.~\ref{eq:diversity}) of their engagement with AI and peer (other student) targets. Two divergent profiles are visible: students with a diverse engagement strategy (non-zero diversity) and students with a focused strategy (zero diversity). Points with identical coordinates are offset horizontally by a small amount. }
    \label{fig:interaction_profiles}
\end{figure}

Our analysis reveals a clear divergence in how students engaged with the resources available in the AI-assisted collaborative environment. As visualized in Figure \ref{fig:interaction_profiles}, student strategies varied along two primary dimensions: the overall quantity of interactions and the evenness with which they were split between the two targets (AI and peers).

\hl{The 27 students produced 258 coded messages over the two weeks, 160 directed to peers and 98 to the AI chatbot. Interaction quantity $s_i$ ranged from 1 to 33 messages (mean 9.6, SD 7.7, with per-student values listed in Appendix}~\ref{app:students}\hl{). }The results show a distinct split in the cohort. A majority of students (n=18) exhibited non-zero diversity in their interactions, flexibly engaging with both their peers and the AI chatbot\hl{ ($D_i$ between 0.49 and 1.00, with twelve of the 18 above 0.75)}. This pattern \hl{indicates that these students drew on both peer and AI resources rather than one alone}. In contrast, a substantial subset of students (n=9) showed a highly focused pattern, interacting exclusively with either the AI or their peers, but not both\hl{. Five students (S1, S3, S4, S10, S19) interacted only with peers and four (S9, S22, S23, S24) only with the AI. The four AI-only students sent 18 messages in total, ten of them Concept/Solution Questions to the chatbot}. This indicates a more specialized, and potentially constrained, approach to the collaborative task.

The quantity and diversity measures together describe student behavior that either one alone would not. For instance, a diversity score of zero indicates a student who may not be developing the ability to use the full range of available resources. \hl{These measures describe the volume and the allocation of a student's interactions. The zero-diversity students indicate an exclusive orientation toward peer or AI. Exclusive orientation toward the AI is a particularly important pattern that may indicate AI reliance, which prior work has linked to mechanisms such as metacognitive laziness }\citep{fan2025beware} and cognitive offloading \citep{yan2024promises,stadler2024cognitive}, \hl{whereas under-utilisation has been associated with trust-calibration issues} \citep{okamura2020adaptive}. Prior work has identified the emergence of AI-centered interaction patterns, in which students tended to interact more with the GAI chatbot than with their group members in AI-assisted collaborative learning \citep{feng2025group}. \hl{Identifying which mechanism underlies a given student's pattern, however, is beyond the scope of this case study. Our aim here is to show that the node-level measures make these patterns visible and quantifiable in the first place.} The node-level analysis here contributes a two-dimensional profile for every student, while one-dimensional measures such as an activity count alone would not separate the nine focused students from the eighteen mixed ones, and a profile built from the AI share alone would not distinguish a student with two messages from one with thirty. \hl{HINA's node-level measures provide process metrics for characterising students' human–AI interaction patterns. These metrics can then be examined alongside outcome performance data, personal attribute data, and qualitative interview data, enabling us to investigate the factors that shape AI usage and to further uncover the mechanisms behind AI usage preferences in GAI-assisted collaborative learning.}

For teachers, the identified interaction profiles enable personalized intervention, allowing them to provide timely support tailored to a student's specific pattern of resource use. For researchers, the individual-level metrics generated by HINA provide a valuable foundation for hypothesis testing. These measures can be correlated with individual attribute data (such as learning gains or survey-based constructs) to investigate the factors influencing interaction propensity. They can also be used as dependent variables to examine differences in engagement patterns across experimental conditions. For instance, researchers can test whether specific instructional designs (e.g., an AI scaffolded to prompt reflection vs. a baseline AI) or different student attributes (e.g., by prior knowledge or attitudes) are associated with statistically significant differences in their engagement quantity and diversity. This approach facilitates direct tests of hypotheses concerning how pedagogical interventions and learner characteristics shape collaborative engagement, which is the kind of evidence theories of learning in human–AI environments require \citep{jarvela2023human}.

\subsubsection{RQ2: Emergent Engagement Patterns}

\begin{figure*}[htbp]
    \centering
    \includegraphics[width=0.8 \textwidth]{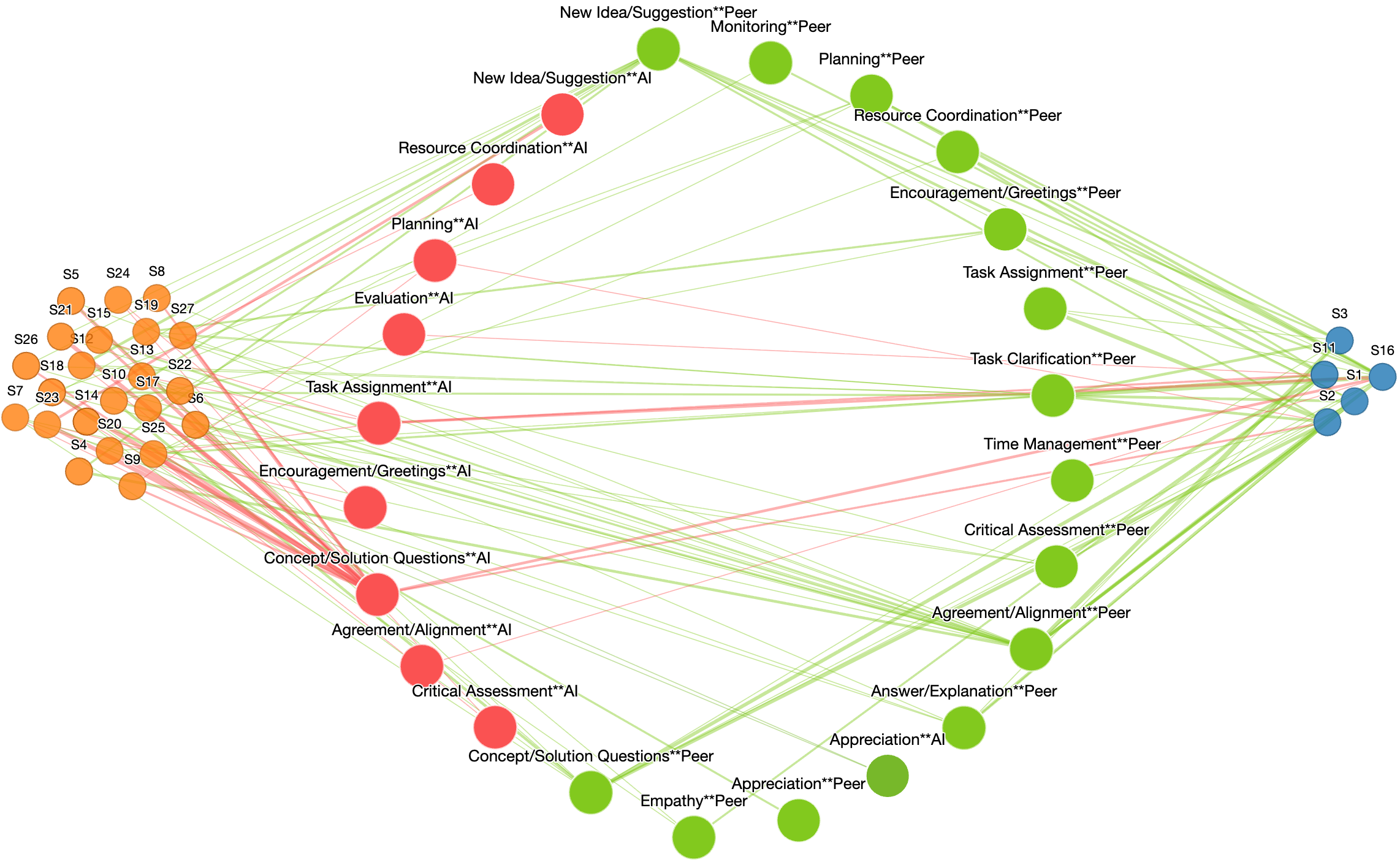}
    \vspace{10pt}
    \caption{Engagement clusters identified through mesoscale analysis of the student-(content code, target) HIN. Student nodes are colored by cluster membership (blue and orange for Clusters 1 and 2 respectively), while (content code, target) nodes are colored in green (content--Peer) and red (content--AI). Double asterisks indicate the composite entities, i.e. ``Appreciation ${}^{\ast\ast}$AI''. \hl{Edges are coloured by the target of each interaction, red for interactions directed at the AI agent and green for interactions with peers, with edge width proportional to the number of interactions. Students in the left cluster (Cluster 2) interact predominantly with the AI agent, as shown by the dense red edges, whereas students in the right cluster (Cluster 1) interact more with their peers, as shown by the predominantly green edges. To present the HINA meso-level clustering result in full, we also provide the detailed partition as an ordered interaction matrix in Appendix}~\ref{app:students}\hl{ (Figure}~\ref{fig:interaction_matrix}\hl{).} The cluster visualization was generated using the HINA web tool at \url{hina-network.com}.}
    \label{fig:engagement_clusters}
\end{figure*}

Based on the clustering results\hl{ (Figure}~\ref{fig:engagement_clusters}\hl{), we identified two distinct engagement patterns among students: Cluster 1 with five students (S1, S2, S3, S11, S16) and Cluster 2 with the remaining 22. The compression ratio of this partition was $\eta=0.967$, and in Appendix}~\ref{app:comparison} \hl{we verified this value against a permutation null model, demonstrating the significant compression obtained by the greedy algorithm. The two clusters differ sharply in how they distributed their interactions, Cluster 1 sending 97 messages to peers and 13 to the AI (an AI share of 12\%), and Cluster 2 sending 63 to peers and 85 to the AI (57\%). Additionally, the two clusters' distributions of interactions over the twelve (code, target) entities with at least five messages differ significantly, with $\chi^2=73.0$ ($df=11$, $p<10^{-10}$). The full student--entity interaction matrix, ordered by cluster, and the relation between the clusters and the RQ1 profiles are given in Appendix}~\ref{app:students}\hl{. }\hl{In Figure}~\ref{fig:engagement_clusters}\hl{, the five Cluster 1 students on the right of the layout connect to the (code, target) entities almost entirely through peer-directed ties (Task Clarification, Agreement/Alignment, Concept/Solution Questions and New Idea/Suggestion with peers), while the 22 Cluster 2 students on the left share the heavily weighted bundle of ties to Concept/Solution Questions directed to the AI. }This result demonstrates that students within each cluster share statistically similar configurations with respect to how they distributed their cognitive, metacognitive, socio-emotional, and coordinative interactions between AI and peer targets. 

\hl{HINA clusters students by their observed interaction behavior first and brings in student characteristics afterwards, rather than splitting the cohort by an attribute (e.g., high- and low-performing students) before the analysis. We did an analysis to examine whether the process similarities captured by the HINA meso-level clusters arise from shared group membership. We examined the association between group and cluster membership using a Fisher–Freeman–Halton exact test on the $2\times 6$ table ($p=0.046$). The test was significant, indicating that cluster membership was associated with group membership, so that students in the same group tended to fall in the same cluster. For larger samples, where the exact computation becomes intractable, a Monte Carlo approximation of the same test can be used instead. This additional analysis demonstrates how HINA's outputs can be related to a categorical learner attribute to understand the mechanisms associated with the emergence of the clustering. Future studies can also use cluster membership as a grouping factor to examine differences across clusters in node-level metrics, prior knowledge, or survey constructs, or combine it with ENA to characterize the interaction patterns of the identified clusters.}

\hl{To demonstrate the necessity and strength of HINA meso-level analysis, we further compared the meso-level clustering method of HINA framework with existing community detection methods. Louvain community detection on the projected student network (Section}~\ref{sec:louvain}\hl{, Appendix}~\ref{app:comparison}\hl{) returns at its default resolution a 15/12 split (adjusted Rand index 0.21--0.22 with the HINA partition across four projection weightings) but does not shorten the description of the HIN ($\eta=0.9999$--$1.0011$). In addition, the number of communities Louvain returns is highly sensitive to a resolution parameter that the analyst must set by hand, ranging from a single community to as many as six to thirteen at other settings, so the "right" number of groups is not determined by the data, as it is possible to determine using HINA. Meanwhile, a K-means analysis requires choosing the number of clusters in advance, which artificially restricts the possible outcomes of the analysis, or employing an arbitrary heuristic (such as the elbow or silhouette method) to select K, to which to results are sensitive. HINA meso-level analysis, by contrast, operates directly on the bipartite structure, without projecting it to a one-mode graph, and clusters only the focal node set, inferring both whether a grouping is warranted and how many groups exist from the data, without a tuning parameter.}

The clustering reveals distinct patterns that future studies could interpret further by examining differences in student metadata (such as academic background, prior experience with AI tools, or learning preferences) between the inferred clusters. Such analysis would help explain why students adopt particular engagement strategies and what factors influence their interaction preferences. The specific characteristics that define each cluster, namely the combinations of interaction codes and targets most representative of each group, are examined in detail through dyadic-level analysis in the following RQ3 results section, where we identify the statistically significant code-target associations that uniquely characterize each cluster's interaction configuration.

\subsubsection{RQ3: Cluster-level Engagement Patterns}

\begin{figure*}[htbp]
    \centering
    \includegraphics[width=\textwidth]{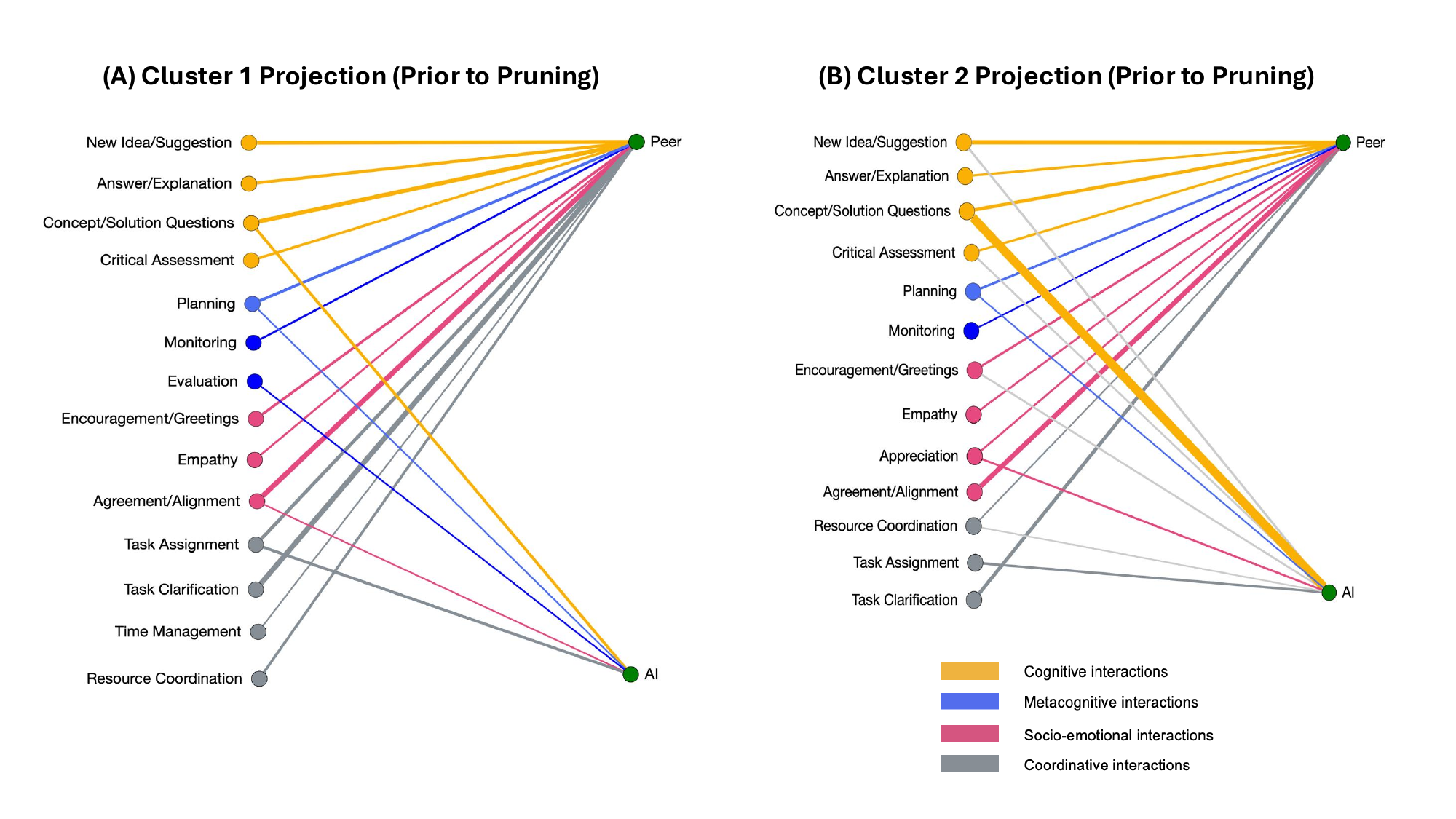}
    \caption{The projected code--target network for each identified cluster, mapping the structural relationships between engagement codes and their associated interaction patterns within each community. The figure gives an initial structural overview of the full distribution of interactions, prior to dyadic-level pruning. Subfigures A and B correspond to Cluster 1 and Cluster 2 respectively; each panel shows all code--target ties, to peers and to the AI, within that cluster.}
    \label{fig:prior_pruning}
\end{figure*}

\begin{figure*}[htbp]
    \centering
    \includegraphics[width=\textwidth]{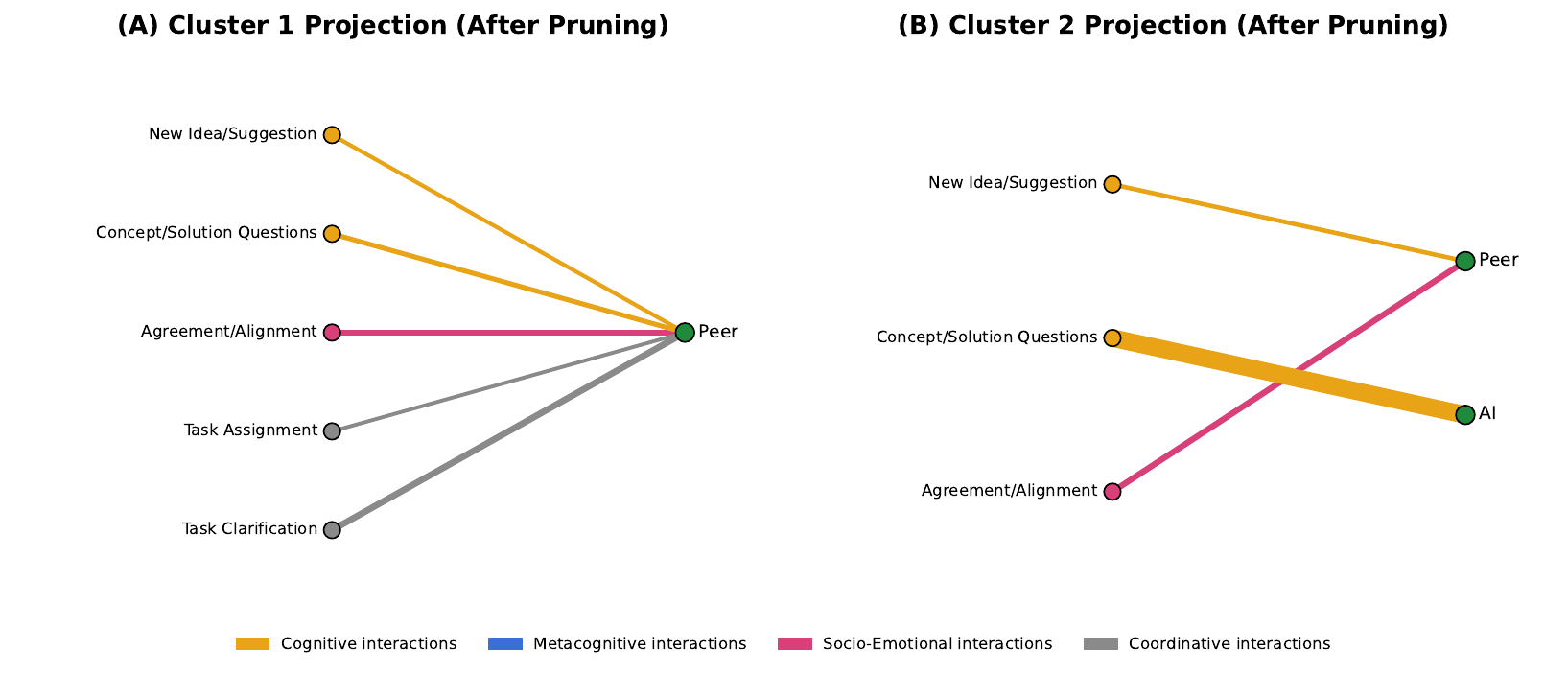}
    \caption{The pruned projected code--target network of each cluster. In this visualization, interactions are color-coded according to their corresponding dimensions: Cognitive, Metacognitive, Coordinative, or Socio-Emotional. Following the dyadic-level network pruning process, only statistically significant associations (at $\alpha=0.05$, default null model with no strengths fixed) are retained. Subfigures A and B correspond to Clusters 1 and 2. The retained edges are listed with their thresholds in Table~\ref{tab:sigedges}.}
    \label{fig:engagement_each_cluster}
\end{figure*}

Based on the RQ3 results shown in Figure \ref{fig:engagement_each_cluster}, the dyadic-level analysis reveals distinct engagement patterns in the two clusters, characterized by different sets of statistically significant code-target associations. Prior to the dyadic-level analysis, the unpruned projected code--target networks shown in Figures \ref{fig:prior_pruning}A and~\ref{fig:prior_pruning}B provide an initial structural overview of the full distribution of interactions within each community. The comparison shows what dyadic-level analysis adds to descriptive visualization: an inferential step that separates significant associations from incidental co-occurrence.

\hl{Table}~\ref{tab:sigedges} \hl{in Appendix}~\ref{app:students} \hl{lists every code--target edge with at least three interactions in each cluster's projected network together with the threshold of the null model, with the retained edges shown in Figure}~\ref{fig:engagement_each_cluster}\hl{. }Cluster 1 (Figure~\ref{fig:engagement_each_cluster}A) exhibits a peer-dominant pattern. This cluster shows significant peer associations \hl{for Cognitive functions (Concept/Solution Questions, 13 interactions; New Idea/Suggestion, 9), a Socio-Emotional function (Agreement/Alignment, 19) and two Coordinative functions (Task Clarification, 20; Task Assignment, 8). Its two largest ties to the AI (five Concept/Solution Questions and five Task Assignments, out of 110 interactions in total) fall below the threshold, so under the null model this cluster's use of the AI is not distinguishable from chance}.

Cluster 2 (Figure~\ref{fig:engagement_each_cluster}B) demonstrates a hybrid engagement pattern with a strong AI component for cognitive information seeking. The most prominent association is between AI and the content code Concept/Solution Questions (71 of the cluster's 148 interactions), indicating that Cluster 2 students primarily used the AI as a resource for resolving queries. This was complemented by peer interactions focused on New Idea/Suggestion (Cognitive, 11) and Agreement/Alignment (Socio-Emotional, 18), suggesting a division of labor in which solution development was primarily handled by the AI while \hl{idea generation and consensus were managed among peers. Two edges (Agreement/Alignment--Peer and New Idea/Suggestion--Peer) are retained in both clusters, with the remaining retained edges specific to an individual cluster (three peer-directed ties in Cluster 1 and Concept/Solution Questions--AI in Cluster 2), distinguishing the two engagement patterns beyond visual inspection.}

The interaction patterns identified for Clusters 1 and 2 reflect different approaches to resource use. By modeling the connections between distinct types of nodes (students, content codes, and targets), HINA shows how different functions are distributed across the learning environment. Cluster 1 showed a \hl{peer-oriented approach, distributing its interactions across cognitive, socio-emotional and coordinative functions among peers, with occasional AI use that does not rise above chance}. Cluster 2, by contrast, employed a targeted division of labor, using AI extensively for concept/solution queries while maintaining peer collaboration for \hl{idea generation and agreement}. \hl{Returning to the common question of how HINA differs from ENA reviewed in Section 2.3, the results of this case study further showcase the distinction. The insights obtained through the HINA  three-level analysis would not have been produced by ENA, which models code co-occurrence within defined units and does not provide individual process metrics, test individual learner–entity ties or infer learner groups from the data. This being said, ENA and HINA can be used together to offer a fuller picture, for instance, with HINA revealing how learners cluster by their interaction patterns and ENA illuminating the epistemic structure of the discourse within each group. }The three-level analysis in HINA quantifies students' human-AI interaction patterns and uncovers their underlying strategies. This capability enables hypothesis testing to identify factors that influence human–AI interaction patterns in collaborative learning contexts, which is a prerequisite for theory development in human–AI collaboration.

\section {Discussion}

\subsection  {HINA: A New Methodological Framework for Modeling Learning Processes}
HINA provides a new learning analytics framework by introducing heterogeneous interaction networks as its foundational modeling paradigm and an integrated multi-level analytical workflow. This lets researchers model learning processes as networks of interactions rather than as descriptive summaries of interactions among various entities such as students, technologies, content, and behavioral constructs. HINA simultaneously provides insights at individual (engagement profiles), dyadic (significant associations), and mesoscale levels (behavioral clusters). This integrated approach addresses the gap in learning analytics by providing quantifiable measures that enable hypothesis testing, data-driven pattern discovery through clustering, and the statistical validation of interaction significance. As a result, HINA's multi-level framework supports both exploratory analysis and theory-building investigations.

What distinguishes HINA is its three-level analytical framework for heterogeneous interactions in learning processes. While HINA shares the data structures used in bipartite and heterogeneous information network analysis \citep{newman2018networks,shi2016survey}, it departs from their typical analytical focuses. Bipartite network analysis often utilizes one-mode projections to collapse the dual-mode structure into a single-entity network to identify clusters or communities of a single node set based on their similarities \citep{latapy2008basic}. In such models, the focus is frequently on partitioning both node sets at once, for example by maximizing a bipartite modularity \citep{guimera2007module}. Heterogeneous information network analysis, for its part, typically prioritizes the calculation of similarity using meta-paths or meta-graphs to determine how ``close'' different types of nodes are to one another for the purposes of recommendation or link prediction \citep{shi2016survey}. These methods often rely on frameworks like PathSim to quantify topological proximity across diverse entity types \citep{sun2011pathsim}, but do not capture the interpretable similarity constructs relevant for the learning context.

In contrast to bipartite and heterogeneous information network analysis, HINA maintains the integrity of the bipartite structure, shifting the analytical focus from node similarity to how nodes are characterized in the particular structural context. For instance, when using HINA to model student-artifact interactions, the framework does not collapse the network to find similar students; instead, it preserves the heterogeneous links to quantify specific engagement patterns and interaction strategies. This functional assessment facilitates the student-centered analysis required in learning analytics research while remaining adaptable to various interaction types in different learning contexts, such as teacher–resource or student-(artifact, behavior). HINA's three-level analytical framework systematically bridges the gap between individual entity profiles (micro-level), dyadic associations (dyadic-level), and engagement clusters (mesoscale). The framework is built on network science methods while making a domain-specific contribution to learning analytics.

\hl{In the case study, we illustrated how the three-level HINA analysis works  to examine human–AI interactions layer by layer. At the node level, the analysis produced learner-level metrics that characterise the intensity and preference of each student's human–AI interactions. At the meso level, the analysis then identified clusters of learners who share similar engagement patterns, inferring both the clusters and their number directly from the data. This separated a small group of highly active, peer-directed students from the rest of the class, whose profiles combined AI-directed questioning with peer exchange. This structure was not specified in advance but emerged from the students' interaction profiles themselves. At the dyadic level, the analysis further characterised how each cluster interacted with peers and the AI differently. Together, these three levels move from individual interaction profiles, to groups of learners with shared patterns, to the specific interaction characteristics that define each group, demonstrating how HINA supports a coherent, progressively deepening analysis of a learning process that no single existing method provides. }Beyond the demonstrated application to human-AI collaboration in the current case study, HINA is applicable to other analytical settings. The framework can be readily adapted to model diverse learning contexts including student interactions with educational technologies, teacher-student-content dynamics in classroom settings, or multimodal engagement patterns in various learning environments. This adaptability positions HINA as a general-purpose framework for learning process analysis, capable of addressing fundamental questions about how different entities in educational ecosystems interact and influence learning outcomes across varied contexts and modalities.

\subsection{Implications for Understanding Human-AI Collaboration}

The application of HINA to human-AI collaborative learning contexts in the case study shows how humans and AI agents distribute cognitive labor in collaborative learning environments, an open question in human-AI collaboration \citep{feng2025mapping, stadler2024cognitive, molenaar2022towards,cukurova2025interplay}. In the case study, the analytical results revealed \hl{that students differed systematically in which kinds of messages they directed to peers and to the AI}. 

Previous studies have identified a GAI-centered interaction pattern \citep{feng2025group} and a spectrum of student-initiated to AI-led inquiries \citep{dang2025human}. The current \hl{case study illustrates how such patterns can be quantified, and is consistent with the past studies in} demonstrating that students exhibit distinct approaches to incorporating AI into their collaborative workflows, with some treating AI as a primary resource for solutions and others maintaining peer-focused strategies with \hl{only occasional} AI use. Understanding these implicit collaboration patterns has significant implications for learning design, as different collaboration approaches may lead to varied learning outcomes, group dynamics, and development of collaborative skills. The ability to identify and quantify these patterns through HINA's analytical framework gives researchers a means of investigating how different collaboration strategies influence educational processes and outcomes.

The results of the case study bear on the design of educational AI systems for promoting human-AI socially shared regulation \citep{jarvela2023human}. Rather than treating AI integration as uniform across students, the case study findings suggest that educational AI should be capable of adapting to students' existing collaboration strategies while also scaffolding the development of more effective approaches. For instance, AI systems might detect when students \hl{rely on the AI to the exclusion of peers} for requesting direct solutions and instead prompt them to engage peers in knowledge construction. Or, alternatively, AI systems may be able to identify when students are under-utilizing their AI resources and suggest specific contexts where AI support would be beneficial. Characterizing human-AI interaction patterns in this way can inform educational AI systems that respond to different learning needs and collaboration styles \citep{giannakos2025promise}.

\subsection {Limitations and Future Research Directions}\label{sec:limitations} 
HINA has several current limitations. Methodologically, HINA's present implementation models heterogeneous interactions in an aggregated manner and does not explicitly capture temporal transitions between interaction states. \hl{Although the case study dataset size is relatively small, its main goal is to demonstrate the analytical steps rather than establish large-scale generalizable patterns of human--AI collaboration in this paper.} For researchers interested in temporal patterns, current workarounds involve segmenting data into phases and conducting separate HINA analyses for each period, then comparing results across phases. This limitation also presents opportunities for methodological advancement. Future development of HINA should consider incorporating dynamic network analysis techniques to model how heterogeneous interactions evolve over time. 

HINA is intended as a general framework for learning process analysis, applicable to any situation in which one aims to understand how learners interact with the tools, artefacts, behaviours or people in their environment. Future work should apply HINA to process data from a range of learning contexts, enabling the identification of generalizable patterns and the development of learning theories in complex social and technological settings. HINA can also be used together with other methods such as process mining, which extracts temporal sequences of targeted behaviors, to combine temporal and relational views of a learning process. Future research should also examine the pedagogical implications of HINA’s analytical results through interviews with students and teachers. This would clarify how HINA’s insights relate to their observations and perceived learning experience, establishing the pedagogical validity of the analytical results in real-world educational contexts. HINA's visualization capabilities also present opportunities for developing interactive teacher dashboards, potentially with generative AI support to improve interpretability.

\section {Conclusion}
This paper has introduced HINA as a methodological framework that models learning processes as heterogeneous interaction networks. By moving beyond homogeneous network models to capture the complex relational ties between students, AI, resources, coded behaviors, and other factors in the learning environment, HINA provides a unified approach to analyzing learning at the individual, dyadic, and mesoscale levels. The case study we present within the AI-assisted collaboration context shows that the HINA framework recovers distinct engagement patterns and strategic roles, and that its outputs can be used for hypothesis-driven research and theory building. HINA allows researchers to move from descriptive results to testable models, and provides a basis for designing learning environments that respond to how students actually distribute their interactions.

\phantomsection
\section*{Acknowledgments} 
\addcontentsline{toc}{section}{Acknowledgments} 

This work is funded by the Hong Kong Research Grant Council under Early Career Scheme \#27605223 (S.F.), and the HKU Institute of Data Science Research Seed Fund (S.F. and A.K).

\section*{Declaration of Conflicting Interest} 

\addcontentsline{toc}{section}{Declaration of Conflicting Interest} 

The author(s) declared no potential conflicts of interest with respect to the research, authorship, and/or publication of this article.

\newcommand{\appsection}[1]{%
  \refstepcounter{section}%
  \addcontentsline{toc}{section}{\protect\numberline{\thesection}#1}%
  \par\bigskip\noindent
  {\large\sffamily\bfseries Appendix \thesection. #1}%
  \par\medskip
}

\appendix
\section{Derivation of the Description Length}\label{app:derivation}

This appendix gives the fixed-length-code derivation of the description length in Eq.~\ref{eq:MDLobj}, following the transmission scheme described in Section~\ref{sec:meso}.

Sparing the technical information theoretic details, in a fixed-length code any quantity $x$ that may take $X$ possible values given the receiver's current knowledge will cost $\log X$ bits to transmit to them. (As before, we use $\log \equiv \log_2$ for brevity.) In our case, we assume that the receiver has minimal knowledge ahead of time: they only know the three constants $\{N_1,N_2,W\}$. We can also in principle include these in the description length, but since they do not impact the final result and are of comparatively little information cost to transmit we can safely ignore them. Given this minimal prior knowledge, we can transmit the node partition $\bm{b}$ and the data $G$ (using this partition) to the receiver, calculating the cost of each step using the method described above.

We can break the information transmission process into three steps: 
\begin{enumerate}

    \item Transmit the community labels $\bm{b}$ of the nodes in $V_1$, by first transmitting the number of groups $B$, then the group sizes $\bm{n}=\{n_r\}_{r=1}^{B}$, then the partition $\bm{b}$. $B$ may take $N_1$ possible values $\{1,...,N_1\}$, and so transmitting $B$ with a fixed-length code costs us $\log N_1$ bits. Then, there are ${N_1-1\choose B-1}$ ways to configure the $B$ non-empty group sizes that constitute $\bm{n}$, and so it takes $\log {N_1-1\choose B-1}$ bits to specify $\bm{n}$. (This can be computed using the classic ``stars and bars'' method from combinatorics.) Finally, given the group sizes $\bm{n}$, there are
    \begin{align}
     ~~~~~~~~~~~~~~~~~~~~~~~~~~~~~~~~~~~~~~~~~~~~~~~~~~~~~~~~~~~~~~~~~~~{N_1\choose n_1,...,n_B} = \frac{N_1!}{\prod_{r=1}^{B}n_r!}
    \end{align}
    ways to configure the partition $\bm{b}$ given the group sizes $\bm{n}$, and the logarithm of this is the number of bits required for transmission of $\bm{b}$ given $\bm{n}$. 

    \item Transmit the total edge weight contributions $w^{(cn)}_{rj}$ from each of the communities $r=1,...,B$ to each of the nodes $j=1,...,N_2$ in the second node set. There are $BN_2$ possible pairs $(r,j)$, and a total weight $W$ to distribute across these pairs, meaning there are ${BN_2+W-1\choose W}$ ways to configure the values $\{w^{(cn)}_{rj}\}$. (We can again use the stars and bars method to compute this quantity.) Thus, the information required to transmit these total community-node weights is $\log {BN_2+W-1\choose W}$.

    \item Transmit $G$ (or, equivalently, the weights $w_{ij}$ of each edge $(i,j)$) given the constraints imposed by the first two steps. There is a total weight of $w^{(cn)}_{rj}$ going from community $r$ in the first node set to node $j$ in the second set, and there are $n_r$ nodes in this community over which we should distribute this weight, so there are ${n_r+w^{(cn)}_{rj}-1\choose w^{(cn)}_{rj}}$ ways to configure the edge weights incident on node $j$ from nodes in community $r$. Since each community $r$ and target node $j$ can be treated independently, there are 
    \begin{align}
    ~~~~~~~~~~~~~~~~~~~~~~~~~~~~~~~~~~~~~~~~~~~~~~~~~~~~~~~~~~~~~~~~~~~~~\prod_{r=1}^{B}\prod_{j=1}^{N_2}{n_r+w^{(cn)}_{rj}-1\choose w^{(cn)}_{rj}}
    \end{align}
    total configurations of the edge weights in $G$ given the previously transmitted constraints. Thus, we require a number of bits equal to the logarithm of this quantity to transmit $G$ after steps (1) and (2) are completed.
    
\end{enumerate}

\section{Comparison with Louvain Community Detection and Permutation Null Models}\label{app:comparison}

\hl{Table}~\ref{tab:louvain} \hl{compares the HINA clustering of the student--(content code, target) HIN (27 students, 24 composite entities, 123 edges, 258 interactions) with Louvain modularity maximization} \citep{blondel2008fast} \hl{applied to four one-mode projections onto the student set: the shared-weight projection, in which the weight between two students is the sum over entities of the smaller of their two interaction counts; the product projection, with similarity $s_{ii'}=\sum_j w_{ij}w_{i'j}$; the cosine-similarity projection used by the HINA web tool; and Newman's collaboration weighting, which divides each shared entity's contribution by the number of students connected to it minus one} \citep{newman2018networks}\hl{. Each projection was analyzed at resolution parameters $\gamma\in\{0.5, 1.0, 1.5\}$ with 20 random seeds. Every resulting partition was scored with the description length of Eq.}~\ref{eq:MDLobj} \hl{and expressed as a compression ratio relative to the trivial partition, so that all partitions are compared on the same scale. Adjusted Rand index (ARI) values are averages over seeds, and the partitions are stable across seeds (mean pairwise ARI between seeds at least 0.90 in every setting).}

\begin{table}[htbp]
\centering
\caption{\hl{Louvain community detection on one-mode projections of the case-study HIN, compared with HINA. $B$: number of communities (range over 20 seeds); ARI: adjusted Rand index against the HINA partition; $\eta$: compression ratio of the partition under Eq.}~\ref{eq:MDLobj}\hl{, averaged over seeds ($\eta>1$ means the partition lengthens the description relative to no partition).}}
\vspace{10pt}
\label{tab:louvain}
\small
\begin{tabular}{l c c l c c}
\hline
Projection & $\gamma$ & $B$ & Sizes (best modularity) & ARI & $\eta$ \\
\hline
all four & 0.5 & 1 & 27 & -- & 1.000 \\
shared weight & 1.0 & 2 & 15, 12 & 0.22 & 0.9999 \\
product & 1.0 & 2 & 15, 12 & 0.21 & 1.001 \\
cosine & 1.0 & 2 & 15, 12 & 0.21 & 1.001 \\
Newman & 1.0 & 2 & 15, 12 & 0.21 & 1.000 \\
shared weight & 1.5 & 7--8 & 8, 6, 4, 4, 2, 1, 1, 1 & 0.02 & 1.233 \\
product & 1.5 & 10--13 & 9, 7, 1, 1, \ldots & $-0.02$ & 1.372 \\
cosine & 1.5 & 12--13 & 10, 5, 2, 1, \ldots & $-0.07$ & 1.398 \\
Newman & 1.5 & 6--8 & 10, 10, 2, 1, 1, 1, 1, 1 & 0.04 & 1.262 \\
\hline
HINA (Eq.~\ref{eq:MDLobj}) & -- & 2 & 22, 5 & 1 & 0.967 \\
\hline
\end{tabular}
\end{table}

\hl{The Louvain partition at $\gamma=1$ separates 12 predominantly peer-oriented students (mean peer share of interactions 0.87) from 15 predominantly AI-oriented students (mean peer share 0.26). This is an interpretable split, but it is one of three qualitatively different answers the method gives depending on $\gamma$, and it does not compress the HIN (the shared-weight projection saves 0.1 bits of the 650.9-bit description, and the other three projections lengthen it).}

\hl{Two permutation null models were used for the compression ratio, each with 200 shuffled networks and the same clustering algorithm applied to each. The first preserves every student's strength $s_i$ and every entity's strength $d_j$ by shuffling the $W=258$ unit-weight interactions among student--entity pairs subject to both constraints (referred to in Sections}~\ref{sec:meso} \hl{and}~\ref{sec:results}\hl{). Under this test the algorithm returned a single cluster in 191 of 200 shuffles, while the compression ratio was $1.000\pm0.002$ (mean $\pm$ SD) with a minimum of 0.977, and the observed value of 0.967 gives $p=0.005$. The second test conducted preserves each student's strength but distributes that student's interactions uniformly at random over the 24 entities. Under this weaker null model the algorithm returned two or more clusters in every shuffle and the compression ratio was $0.981\pm0.006$ with a minimum of 0.966, so the observed value corresponds to $p=0.015$. The difference between the two null models is informative, since when students differ only in activity level, the description length objective already separates high-activity from low-activity students, so activity level is part of what defines a cluster under Eq.}~\ref{eq:MDLobj}\hl{. The result under the first null model shows that the case-study clusters capture structure beyond both activity level and entity popularity, while the second null model shows that activity heterogeneity alone produces compression ratios much closer to the observed value (the observed value lies near the lower end of that null distribution rather than far outside it), so the stronger evidence for the clusters comes from the joint null model. The distribution of $\eta$ under the first null model falls exactly at 1.000 in 191 shuffles with only nine smaller values, which is why its standard deviation is so small relative to the distance to its minimum.}

\section{Additional Case-Study Results}\label{app:students}

\begin{figure*}[htbp]
    \centering
    \includegraphics[width=0.9\textwidth]{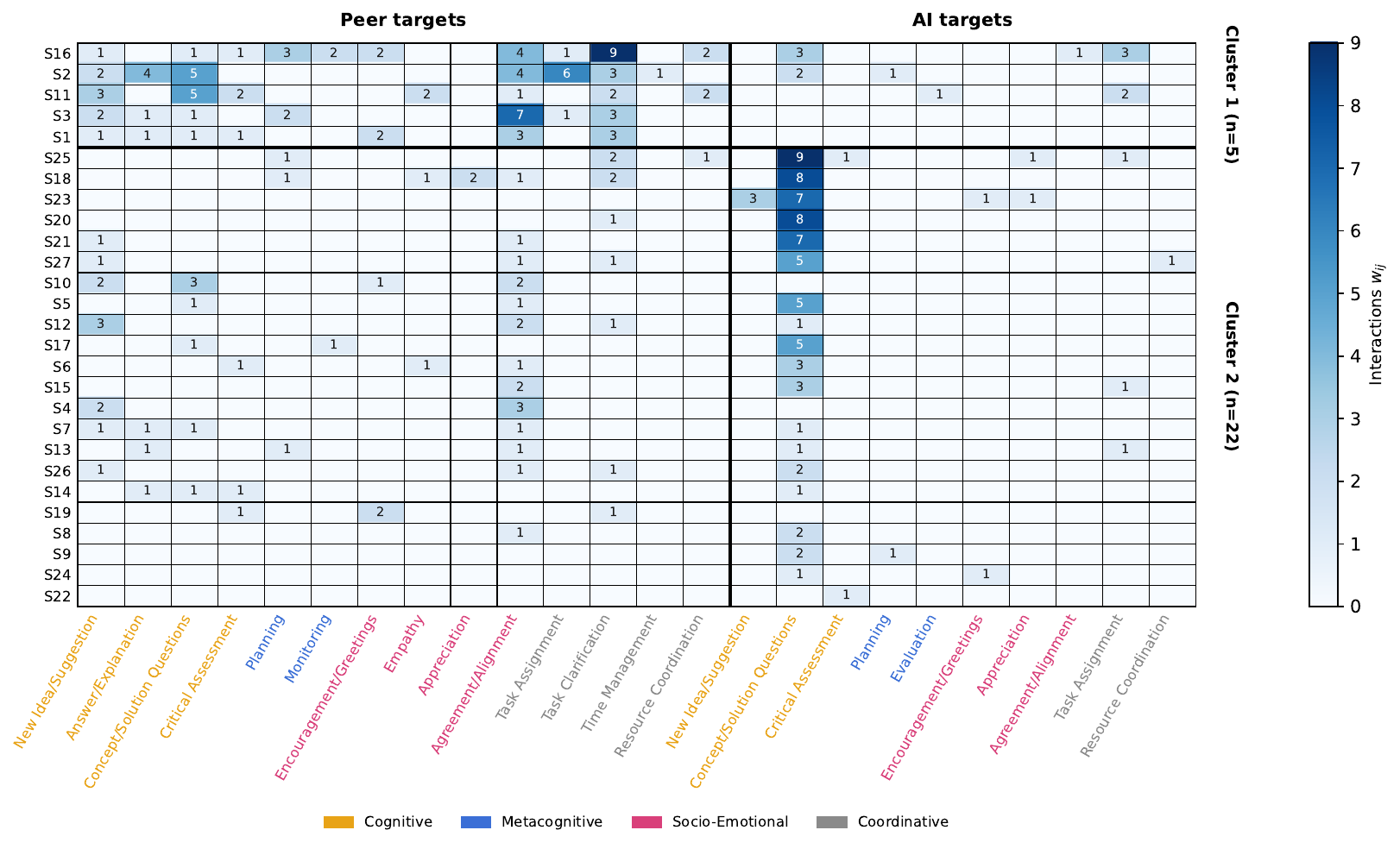}
    \caption{\hl{Engagement clusters of the student-(content code, target) HIN, shown as the matrix of interaction counts $w_{ij}$ between each student (rows) and each (content code, target) entity (columns). Rows are grouped by inferred cluster (Cluster 1: five students; Cluster 2: 22 students) and ordered by interaction quantity within each cluster. Columns are grouped by target (peer or AI) and colored by the dimension of the content code. Cluster 1 concentrates on peer-directed coordinative, socio-emotional and cognitive interactions. Meanwhile, Cluster 2 concentrates on Concept/Solution Questions directed to the AI. The same clusters can be viewed as an interactive network in the HINA web tool at} \url{hina-network.com}\hl{.}}
    \label{fig:interaction_matrix}
\end{figure*}

\hl{Figure}~\ref{fig:interaction_matrix} \hl{shows the full student--entity interaction matrix ordered by cluster, complementing the network view in Figure}~\ref{fig:engagement_clusters}\hl{. }\hl{Cluster 1 consists of the five students with the most peer-directed interactions (12 to 26 peer messages, compared with at most 8 in Cluster 2, with overall $s_i$ between 12 and 33, compared with 1 to 16 in Cluster 2). In terms of the RQ1 profiles it contains two of the five peer-only students (S1, S3) and three high-volume mixed students, while all four AI-only students fall in Cluster 2. This forms a link between the two analyses, since RQ1 categorizes each student on the coarse student--target HIN, and RQ2 refines the target into (code, target) entities so that the clustering can identify what was directed at whom. Because the description length depends on interaction counts, the quantity dimension of the RQ1 profile carries over into the clustering.}

\hl{Table}~\ref{tab:sigedges} \hl{lists the dyadic-level analysis of the two cluster-level code--target networks underlying Figure}~\ref{fig:engagement_each_cluster}\hl{. For each cluster, all code--target edges with at least three interactions are listed with their weight, and edges are retained if the weight strictly exceeds the $95\%$ quantile of the Binomial null model with no strengths fixed (Section}~\ref{sec:dyadic}\hl{), which is $q=7$ for Cluster 1 ($W=110$ interactions, 14 codes $\times$ 2 targets) and $q=10$ for Cluster 2 ($W=148$, 13 codes $\times$ 2 targets). The number of codes is the number observed in each cluster. The null probability of a weight at least as large as the smallest retained weight is 0.044 (Cluster 1) and 0.028 (Cluster 2).}

\begin{table*}[!t]
\centering
\caption{\hl{Dyadic-level descriptive results for the (Code, Target) HINs characterizing the two identified clusters}}
\label{tab:sigedges}
\small
\vspace{10pt}
\begin{tabular*}{\textwidth}{@{\extracolsep{\fill}} l l r c | l l r c}
\hline
\multicolumn{4}{c|}{\textbf{Cluster 1} ($n=5$; threshold $w>7$)} & \multicolumn{4}{c}{\textbf{Cluster 2} ($n=22$; threshold $w>10$)} \\
Code & Target & $w$ & Retained & Code & Target & $w$ & Retained \\
\hline
Task Clarification & Peer & 20 & yes & Concept/Solution Questions & AI & 71 & yes \\
Agreement/Alignment & Peer & 19 & yes & Agreement/Alignment & Peer & 18 & yes \\
Concept/Solution Questions & Peer & 13 & yes & New Idea/Suggestion & Peer & 11 & yes \\
New Idea/Suggestion & Peer & 9 & yes & Task Clarification & Peer & 9 & no \\
Task Assignment & Peer & 8 & yes & Concept/Solution Questions & Peer & 7 & no \\
Answer/Explanation & Peer & 6 & no & Answer/Explanation & Peer & 3 & no \\
Concept/Solution Questions & AI & 5 & no & Critical Assessment & Peer & 3 & no \\
Planning & Peer & 5 & no & Encouragement/Greetings & Peer & 3 & no \\
Task Assignment & AI & 5 & no & New Idea/Suggestion & AI & 3 & no \\
Critical Assessment & Peer & 4 & no & Planning & Peer & 3 & no \\
Encouragement/Greetings & Peer & 4 & no & Task Assignment & AI & 3 & no \\
Resource Coordination & Peer & 4 & no & & & & \\
\hline
\end{tabular*}
\end{table*}

\phantomsection

\clearpage 
\balance


\end{document}